\documentclass[preprint,12pt]{elsarticle}

\usepackage{epsfig}

\usepackage{amsmath}
\usepackage{amsfonts}
\usepackage{amssymb}
\usepackage[amssymb]{SIunits}
\usepackage{algorithm}
\usepackage{algorithmic}
\usepackage[english,linesnumbered,algo2e,algoruled,vlined]{algorithm2e}
\usepackage[dvips]{color}
\usepackage{pstricks}
\usepackage{pst-tree}
\usepackage{pst-node}

\usepackage{lscape}
\usepackage{multirow}

\usepackage{lscape}
\usepackage{graphicx}
\usepackage{subfigure}
\usepackage{verbatim}

\newtheorem{thm}{Theorem}[section]

\newtheorem{prop}[thm]{Proposition}

\journal{Artificial Intelligence}

\begin{document}

\begin{frontmatter}

\title{A Formal Framework for Mobile Robot Patrolling in Arbitrary Environments with Adversaries}

\author{Nicola Basilico}  \author{Nicola Gatti\corref{ngatti}} \author{Francesco Amigoni} 

\address{Artificial Intelligence and Robotics Laboratory \\ Dipartimento di Elettronica e Informazione, Politecnico di Milano \\ Piazza Leonardo da Vinci 32, 20133 Milano, Italy}

\cortext[ngatti]{Corresponding author.  Voice: +39 02 2399 3658. Fax: +39 02 2399 3411. Email: ngatti@elet.polimi.it.}

\maketitle

\begin{abstract}
Using mobile robots for autonomous patrolling of environments to prevent intrusions is a topic of increasing practical relevance. One of the most challenging scientific issues is the problem of finding effective \emph{patrolling strategies} that, at each time point, determine the next moves of the patrollers in order to maximize some objective function. In the very last years this problem has been addressed in a game theoretical fashion, explicitly considering the presence of an adversarial intruder. The general idea is that of modeling a patrolling situation as a game, played by the patrollers and the intruder, and of studying the equilibria of this game to derive effective patrolling strategies. In this paper we present a game theoretical formal framework for the determination of effective patrolling strategies that extends the previous proposals appeared in the literature, by considering environments with arbitrary topology and arbitrary preferences for the agents. The main original contributions of this paper are the formulation of the patrolling game for generic graph environments, an algorithm for finding a deterministic equilibrium strategy, which is a fixed path through the vertices of the graph,  and an algorithm for finding a non-deterministic equilibrium strategy, which is a set of probabilities for moving between adjacent vertices of the graph. Both the algorithms are analytically studied and experimentally validated, to assess their properties and efficiency.
\end{abstract}

\end{frontmatter}

\section{Introduction}
\label{section:introduction}

A patrolling situation is characterized by one or more \textit{patrollers} and by some \textit{targets}, located in an environment, that have to be patrolled in order to prevent the entering of a possible \emph{intruder}.  Patrollers move in the environment according to a \emph{patrolling strategy} that, at each time point, determines their next action. One of the most interesting scientific issues is the determination of effective patrolling strategies for autonomous robotic patrollers, namely, of strategies that drive the robots around environments to maximize some objective function. Several objective functions can be defined according to the specific patrolling situation, and works in literature can be approximately divided into two main classes. \textcolor{black}{In the first one, the presence of the adversary, i.e., the intruder, is not taken into account and the problem of determining the optimal patrolling strategy comes down to the problem of exploring or covering the environment. This problem can be formulated as a classic decision theory problem where the objective function can be, for instance, a coverage index or an entropy-based metric~\cite{yu2004,Sujan2005,rocha2005,Stachniss,sim}.} In the second class of works, the presence of an adversary is considered and the problem is addressed in a game theoretical fashion~\cite{amigoni-iat2008,GalKaminkaICRA2008,paruchuri2008,paruchuri2007}. In this case, the patrollers'  objective function to be maximized is their expected utility (in the sense of von~Neumann and Morgenstern~\cite{fudenberg1991}) computed over the game outcomes. \textcolor{black}{The intruder is usually modeled as a rational agent that observes the strategy of the patrollers and acts to maximize its expected utility. That is, the intruder is assumed to be as strong as possible.} Roughly speaking, the contributions belonging to the first class come mostly from the mobile robotic community, while the contributions belonging to the second class come mostly from the multiagent community. Recently, it has been shown that considering a model of the adversary can give the patrolling robots a larger expected utility than the case in which the opponent is not modeled~\cite{amigoni-iat2008}. 

In this paper we present a formal framework for the determination of effective patrolling strategies  that belongs to the class of game theoretical approaches. The proposed approach is based on the general idea of modeling a patrolling situation as a game, played by the patrollers and the intruder, and of studying the \emph{equilibria}~\cite{fudenberg1991} of this game to derive the optimal patrolling strategy. Now we survey the main related works, to set the background for introducing in more detail our original contributions.

\subsection{Game Theoretical Patrolling Models for Mobile Robots}
\label{subsection:relatedworksmodel}

Game theoretical approaches model a patrolling situation as a \emph{non-cooperative game}~\cite{fudenberg1991} between the patrollers and the intruder. Two are the main approaches proposed in the literature for robotic patrolling with adversaries: one does not explicitly model the preferences of the adversaries~\cite{GalKaminkaICRA2008}, whereas the other one does~\cite{paruchuri2008,paruchuri2007}. Before reviewing these approaches, we note that similar strategic problems have been addressed in the pursuit-evasion field (e.g.,~\cite{isler200tro,vidal2002tra}) and by hide-and-seek approaches~\cite{ConitzerIJCAI2009}, where a hider can hide itself in a vertex of an arbitrary graph and a seeker can move along the graph to seek the hider within a finite time. \textcolor{black}{However, some assumptions, including the fact that the evader's goal is only to avoid capture and not to enter an area of interest, make the pursuit-evasion problem not directly comparable with the patrolling problem we are considering.}

\textcolor{black}{We now describe the main approach that does not model the intruder preferences.} In~\cite{GalKaminkaICRA2008}, the authors consider the problem of patrolling a perimeter divided in cells, each one giving access to an area of interest, by employing a team of synchronized mobile robots acting in turns. The perimeter is considered as a ring whose cells require the same time, say $d$ turns, to the intruder for entering. The robots keep an evenly separated formation by moving in a coordinated fashion. (The authors show that this patrolling configuration is optimal for their settings.) The patrolling strategy does not depend on the specific cells in which the robots are. The authors present different movement models for the robots. In the simplest one, all the robots move clockwise with probability $p$ or move counterclockwise with probability $1-p$. In the most realistic movement model, all the patrollers move  to the cells they are headed to with probability $p$ and reverse their heading, staying in their current cells for a turn, with probability $1-p$. \textcolor{black}{The intruder is assumed to be in the position to repeatedly observe the actions of the patroller (staying hidden), derive a correct belief over the patroller's strategy, and attack the cell for which the probability to be captured is the smallest one (this is because no preferences over the cells are considered).} The optimal patrolling strategy amounts to choose the value of $p$ that maximizes the minimum expected utility for the patrollers or, equivalently, that max-minimizes the detection probability. Two interesting extensions to this work are worth citing. In~\cite{agmonTipi}, \textcolor{black}{the authors study the impact of the intruder's knowledge on the performance of the patrolling strategy (e.g., when the intruder has zero or partial knowledge over the patroller's strategy).} In~\cite{kaminkaAAMAS2009}, \textcolor{black}{the authors study the impact  of uncertainty over the sensed data.} These works present two main limitations. First, they are applicable only to very special ring-like environments where all the cells have the same penetration time and the patrollers have no preferences over the cells. Second, the strategy they produce is optimal only when the intruder has no preferences over the cells. The work we present in this paper overcomes these limitations by considering arbitrary topologies and preferences for patroller and intruder.

\textcolor{black}{We now turn to describe the main approach that explicitly models the intruder's preferences.} In~\cite{paruchuri2007}, the authors deal with the problem of patrolling $n$ areas by using a single robotic patroller such that the number of turns it would spend to patrol all the areas is strictly larger than time $d$ needed by the intruder to enter an area. They model such a problem as a two-player (i.e., the patroller and the intruder) strategic-form game with incomplete information (i.e., the intruder's preferences over the areas can be uncertain to the patroller)~\cite{fudenberg1991}. The actions available to the patroller are all the possible routes of $n$ areas, while the intruder chooses a single area to enter. The intruder is assumed to be in the position to repeatedly observe the actions of the patroller (staying hidden), derive a correct belief over the patroller's strategy, and find its best response to the patroller's strategy. The appropriate equilibrium concept, in which the patroller maximizes its expected utility, is the \emph{leader-follower} equilibrium~\cite{commitmentbased}. (A slight variation of this approach has been applied to the problem of patrolling $n$ access points with $m<n$ static checkpoints at the Los Angeles International Airport~\cite{armor}.) As discussed in~\cite{GattiECAI2008}, the approach in~\cite{paruchuri2007} presents two drawbacks. First, no topology connecting the areas is considered and therefore it can hardly be applied to real-world patrolling settings, where areas are usually connected by intricate topologies and the patroller cannot move between any two areas in a single turn. Second, if the decisions of the patroller are over the \emph{next route} to patrol (instead of over the next area), the intruder can increase its expected utility by waiting, observing the patroller's actions, and then choosing the turn in which to enter. The work we present in this paper eliminates these drawbacks, by considering settings with arbitrary topologies and allowing the patroller to decide over the next area to patrol.

\subsection{Main Original Contributions}

We present an approach to robotic patrolling that is more general than the game theoretical approaches discussed above, since it deals with environments with arbitrary topology and with arbitrary preferences for the agents. In particular, the main original contributions of this paper can be summarized as follows.
\begin{itemize} 
\item We formulate the patrolling problem as a game played by the patroller and an intruder in an environment described as a graph. In this setting, the patrolling strategy amounts to determine the vertex the patroller should visit next. Assuming that the intruder can observe the patroller for an indefinitely long time, the optimal patrolling strategy is found by calculating the leader-follower equilibrium of the game. 
\item We propose an algorithm to find a deterministic equilibrium strategy, which consists in a fixed path through the vertices of the graph such that, when the patroller follows it, attempting an intrusion is not the best action for a rational intruder. 
\item We propose an algorithm to find a non-deterministic equilibrium strategy, which consists in a  set of probabilities for moving between adjacent vertices of the graph such that, when the patroller follows it, its expected utility is maximized.
\end{itemize}
Both the algorithms are analytically studied and experimentally validated, to assess their properties and efficiency. We explicitly note that some preliminary results about the algorithms for finding the deterministic and non-deterministic equilibrium strategies have been reported in~\cite{BasilicoGattiAmigoniIAT2009} and in~\cite{BasilicoGattiAmigoniAAMAS2009,AmigoniBasilicoGattiICRA2009,BasilicoGattiRossiCIG2009,BasilicoGattiRossiCeppiAmigoniIAT2009}, respectively.

\subsection{Structure of the Paper}

In the next section we formulate the problem and we overview the main results presented in the rest of the paper. In Sections~\ref{S:deterministic} and~\ref{section:nondeterministic} we present the algorithms for finding the deterministic and the non-deterministic equilibrium strategies, respectively. Both these sections have a similar structure, presenting the specific state of the art, introducing the proposed algorithms, analyzing them theoretically, and validating them experimentally. Section~\ref{S:extensions} discusses some extensions to our framework in order to capture more realistic aspects and to improve its efficiency. Finally, Section~\ref{S:conclusions} concludes the paper.

\section{Problem Formulation and Overview of the Main Results}
\label{section:statement}
In this section, we formalize the patrolling setting we study (Section~\ref{subsection:problemstatement}), we present our game model (Section~\ref{subsez:gamemodel}), we discuss the appropriate solution concept (Section~\ref{subsez:stackelbger}), and we summarize the main results we provide in this paper (Section~\ref{subsection:summaryofresults}).

\subsection{Patrolling Setting}
\label{subsection:problemstatement}

\begin{figure}[h]
\centering
\includegraphics[scale=0.65]{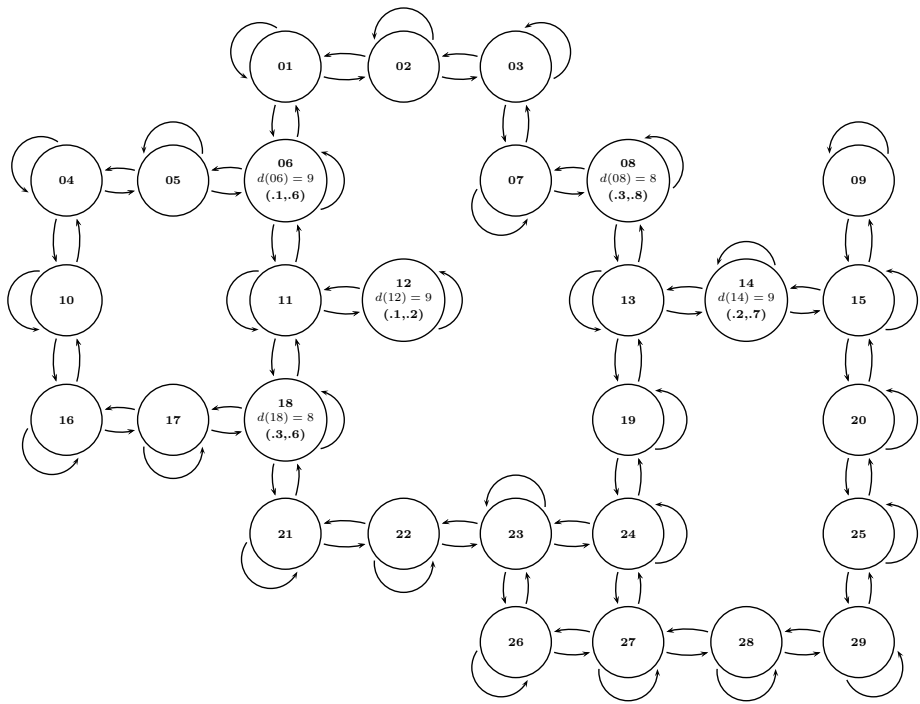}
\caption{The graph representing the patrolling setting used as running example.}
\label{F:graforunningexample}
\end{figure}

\begin{figure}
\centering
\includegraphics[scale=0.6]{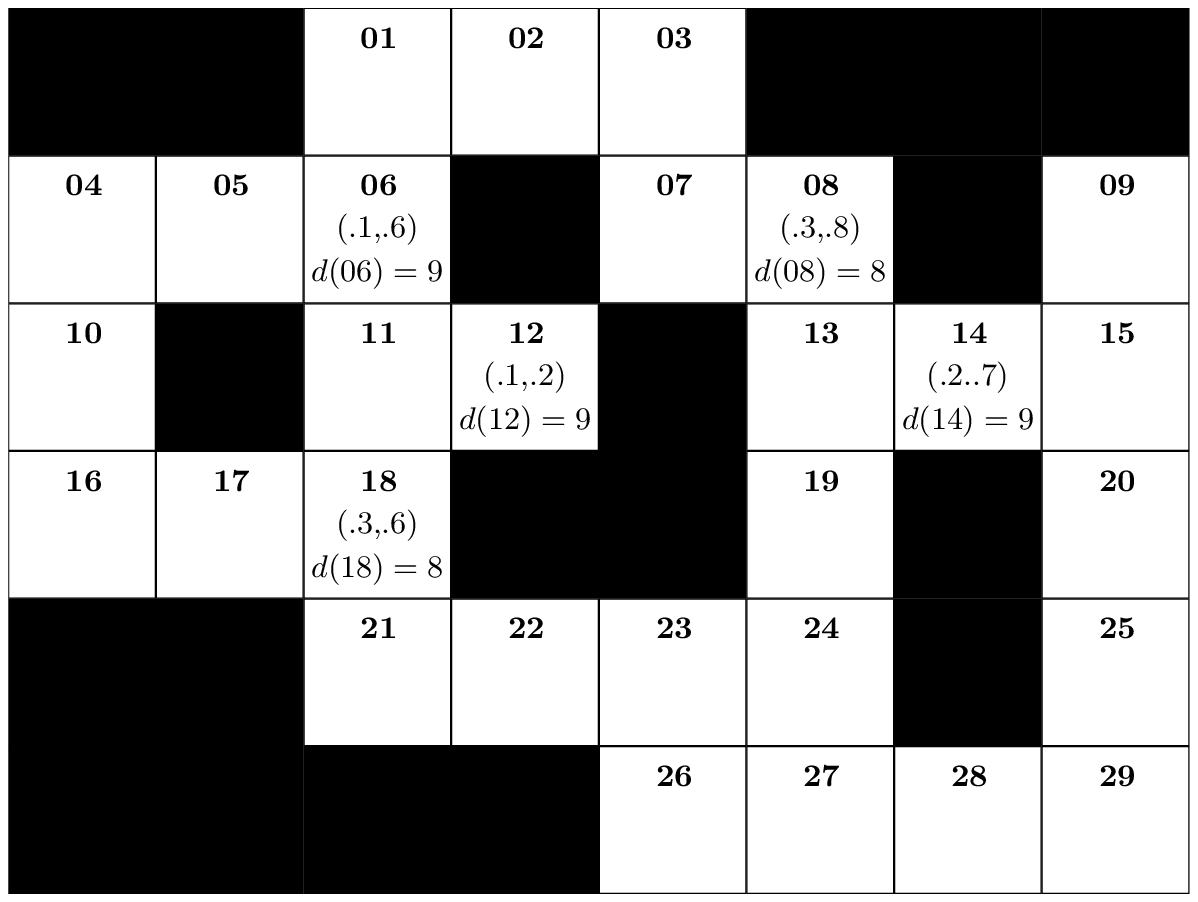}
\caption{Grid map of the environment corresponding to the graph of Figure~\ref{F:graforunningexample}.}
\label{F:mapparunningexample}
\end{figure}

We study settings that are characterized by the following features:
\begin{itemize}
\item the environment is represented by a directed graph (as in~\cite{GalKaminkaICRA2008});
\item there is a single patrolling robot equipped with sensors (e.g., a camera) able to detect intruders (as in~\cite{paruchuri2008,paruchuri2007});
\item the intruder can perfectly observe the patroller's strategy before acting and derive a correct belief over it (as in~\cite{GalKaminkaICRA2008,paruchuri2007});
\item time is discretized in turns (as in~\cite{GalKaminkaICRA2008,paruchuri2007});
\item the intruder enters in vertices and cannot do anything else for some turns once it has attempted to enter a vertex (this amounts to say that penetration takes some turns to be completed, as in~\cite{GalKaminkaICRA2008,paruchuri2007});
\item the patroller and the intruder are assumed to be rational agents (as in~\cite{paruchuri2008,paruchuri2007}).
\end{itemize}
The patrolling setting is described by a direct graph $G=(V,A,T, v,d)$. $V$~is a set of $n$~vertices to be patrolled. $A$~is the set of arcs connecting the vertices.  We often represent $A$~by a function $a:V\times V\rightarrow \{0,1\}$, where $a(i,j)=1$ means that there exists an arc directed from vertex~$i$ to vertex~$j$ and $a(i,j)=0$ means that there is not. Given a vertex~$i$, a vertex~$j$ is \emph{adjacent} to~$i$ if $a(i,j)=1$. $T\subseteq V$ contains the vertices that have some value for both the patroller and the intruder. We call these vertices \emph{targets}. In practical applications, a target may represent an access point to an area with some value (e.g., a door, as in~\cite{GalKaminkaICRA2008}) or an area with some value (e.g., an house, as in~\cite{paruchuri2007}). Vertices that are not targets (in $V \setminus T$) are part of paths that the patroller traverses to move between targets. $v$ is a pair of functions $\{v_{\mathbf{p}},v_{\mathbf{i}}\}$ where $v_{\mathbf{p}}:T\rightarrow \mathbb{R}$ assigns each target a value for the patroller and $v_{\mathbf{i}}:T\rightarrow \mathbb{R}$ assigns each target a value for the intruder. Patroller and intruder can assign different values to the same target. The function $d:T\rightarrow \mathbb{N}\setminus \{0\}$ assigns each target a time interval (measured in turns) that the intruder must spend to successfully enter it. We call $d(i)$ the intruder's \emph{penetration time} for target~$i$. We discuss in Section~5 how we can deal with situations wherein the values of~$d(\cdot)$ and~$v_{\mathbf{i}}(\cdot)$ are uncertain.  An example of a patrolling setting captured by our model is shown in Figure~\ref{F:graforunningexample}. The bold numbers identify the vertices; arcs are depicted as arrows; the set of targets is $T=\{06,08,12,14,18\}$; the values reported in target~$i$ are~$d(i)$ and~$(v_{\mathbf{p}}(i),v_{\mathbf{i}}(i))$. The graph representation of an environment can be derived from a grid map  of the environment, for example as discussed in~\cite{kolling2008}.  The graph of Figure~\ref{F:graforunningexample} can represent the grid environment of Figure~\ref{F:mapparunningexample}, where every white cell of the grid corresponds to a node in the graph (black cells are obstacles). In the following, we complete the description of our setting by introducing sensing and action capabilities of the patroller and of the intruder.

The sensing capabilities of the patroller are defined by a function $S:V\times V\rightarrow [0,1]$ where $S(i,j)$ is the probability with which the patroller, given that its current vertex is~$i$, detects an intruder that is in vertex~$j$. In Sections~3 and~4, we assume that the patroller can sense only its current vertex without any uncertainty. Formally, we assume that $S(i,j)=1$ only if~$i=j$, and $S(i,j)=0$ otherwise. In Section~5 we discuss how our results can be extended to the general case in which patroller's perceptions are described by a generic $S(\cdot,\cdot)$. 
When the patroller detects the intruder, we say also that the patroller \emph{captures} the intruder. The intruder is in the position to observe the movements of the patroller along the graph and to derive a correct belief over the patroller's strategy. That is, the intruder knows the patroller's strategy before acting. In Section~5 we discuss how our results can be extended to the case in which the intruder cannot perfectly observe the environment before acting.

Time is discretized in turns and $k\in \mathbb{N}$ denotes a turn. We assume a simple movement model for the patroller: it spends one turn to move between two adjacent vertices in~$G$ and patrol the arrival vertex. As in~\cite{GalKaminkaICRA2008,paruchuri2008}, we assume that the intruder is able to appear directly in a target when it decides to enter and to disappear directly from the entered target. We recall that, when an intrusion is attempted in target $t$, the intruder stays there and cannot do anything else for $d(t)$ turns. During these turns the intruder can be detected (captured) by the patroller.  We discuss in Section~5 how the above model can be extended to situations in which the intruder reaches targets by moving along paths and there is a delay between the turn at which the intruder makes the decision to enter a target and the turn at which it actually enters.

\subsection{Game Model}
\label{subsez:gamemodel}
The game model we employ to capture a patrolling problem belongs to the class of two-player extensive-form games with imperfect information. In particular, we use a two-player dynamic repeated game~\cite{fudenberg1991}, where the players are the patroller agent and the intruder agent. (The game can be represented also as a partially-observable stochastic game with infinite states. However, since this representation does not provide any advantage, we do not discuss it.) The game develops in turns.  At each turn, a strategic-form game is played in which the players act simultaneously. The patroller chooses the next vertex to reach among those adjacent (directly connected) to its current vertex, formally, called $i$ the vertex of the patroller at turn $k$, its available actions are $\textit{move}(j)$, for all vertices $j$, such that $a(i,j)=1$. At turn $k$, the intruder, if it has not previously attempted to enter any target, chooses whether or not to enter a target and, in the first case, what target to enter, formally, its actions are $\textit{wait}$ and $\textit{enter}(i)$. If, instead, the intruder has previously attempted to enter a target $i$, it cannot take any action for $d(i)$ turns after having attempted to enter. This repeated game is dynamic since it changes at each turn: the positions of the patroller (i.e., its current vertex) and of the intruder (i.e., attacking a target or waiting) change. The game is with imperfect information since, when the patroller acts, it does not know whether the intruder is currently within a vertex or it is still waiting to attack. That is, the intruder's actions are not perfectly observable. The game has an infinite horizon, since the intruder is allowed to wait indefinitely outside the environment.

\begin{figure}[h]
\centering
\includegraphics[scale=.9]{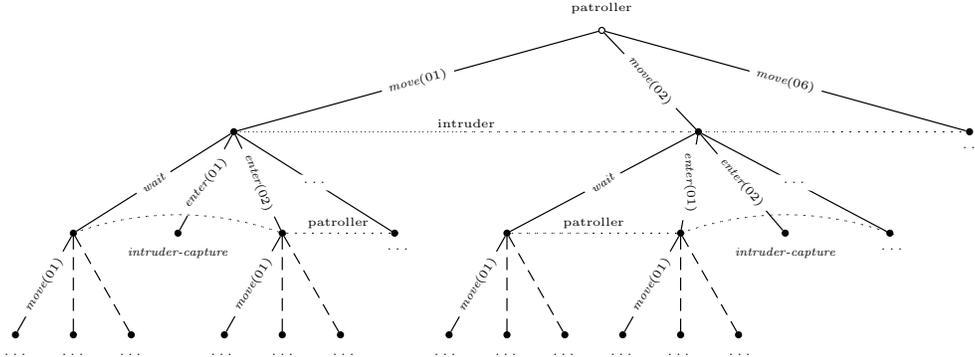}
\caption{A portion of the game tree for the patrolling setting of Figure~\ref{F:graforunningexample}, with the patroller initially in $01$.}
\label{F:tree}
\end{figure}

Figure~\ref{F:tree} reports a portion of the extensive-form representation of the patrolling game for the setting of Figure~\ref{F:graforunningexample}, given that the initial position of the patroller is vertex $01$. Branches represent actions and players' information sets are depicted as dotted lines. We recall that an information set of a player is a set of decision nodes of the player that it cannot distinguish~\cite{fudenberg1991}. That is, when the player is in any decision node of the information set, it just knows to be in a node of that set, but it does not know in which specific node. Information sets are used to represent players' imperfect observation over the actions of their opponents. In our game tree, we use information sets in two ways. First, we use information sets of the intruder: given a decision node $\eta$ of the patroller, all the decision nodes of the intruder that are direct descendent of $\eta$ constitute an information set. This is in accordance with the fact that, at each turn, the players act simultaneously, and the the intruder cannot observe the last action undertaken by the patroller. Second, we use information sets of the patroller to represent the fact that it cannot observe the intruder's actions except when it detects the intruder in some vertex.

The possible outcomes of the game are:
\begin{itemize}
\item \emph{no-attack}: when the intruder never enters any target;
\item \emph{intruder-capture}:  when the intruder attempts to enter a target~$i$ at turn~$k$ and the patroller detects the intruder in target~$i$ in the time interval $\{k,k+1,\ldots,k+d(i) - 1\}$;
\item \emph{penetration-$i$}:  when the intruder enters a target~$i$ at turn~$k$ and the patroller does not detect the intruder in target~$i$ in the time interval $\{k,k+1,\ldots,k+d(i) - 1\}$.
\end{itemize}
Agents' utility functions over the outcomes are defined as follows. The patroller's utility function, denoted by $u_{\mathbf{p}}$, depends on the values $v_{\mathbf{p}}$ of the preserved targets. We assume the patroller to be risk neutral, and therefore:
\[
u_{\mathbf{p}}(x)=
\begin{cases}
\sum_{i\in T}v_{\mathbf{p}}(i) & x=\textit{no-attack} \textnormal{ or  } \textit{intruder-capture}\\
\sum_{i\in T,i \neq j}v_{\mathbf{p}}(i) & x=\textit{penetration-}j \\
\end{cases}.
\]
Notice that the patroller gets the same utility when the intruder is captured and when the intruder never enters. This is because, in the case a utility surplus is given for capture, the patroller could prefer a lottery between $\textit{intruder-capture}$ and $\textit{penetration-}i$ to $\textit{no-attack}$. This behavior is not reasonable, since the patroller's purpose is to preserve as much value as it can. In the case the patroller is risk averse, $u_{\mathbf{p}}$ should be defined as a concave function of the sum of the preserved values. In the case it is risk seeking, $u_{\mathbf{p}}$ should be defined as a convex function. 

The intruder's utility function, denoted by $u_{\mathbf{i}}$, depends on the value $v_{\mathbf{i}}$ of the attacked target. We assume the intruder to be risk neutral, and therefore:
\[
u_{\mathbf{i}}(x)=
\begin{cases}
0 & x=\textit{no-attack} \\
-\epsilon & x= \textit{intruder-capture}\\
v_{\mathbf{i}}(i) & x=\textit{penetration-}i \\
\end{cases},
\]
where $\epsilon\in \mathbb{R}^+$ is a penalty due to the capture. This is to say that the \emph{status quo} (i.e., $\textit{no-attack}$) is better than being captured (i.e., $\textit{intruder-capture}$) for the intruder. In the case of risk averse or risk seeking intruder, $u_{\mathbf{i}}$ should be concave or convex, respectively.

Formally, we define the space $H$ of all the possible histories $h$s of vertices visited (or, equivalently, actions taken) by the patroller. For example, for the setting of Figure~\ref{F:graforunningexample}, given that the patroller starts from vertex $01$, a possible history is $h=\langle 01, 02, 03, 07, 08 \rangle$. We define the \emph{patroller's strategy} as $\sigma_{\mathbf{p}}:H\rightarrow \Delta(V)$ where $\Delta(V)$ is a probability distribution over the vertices $V$ or, equivalently, over the corresponding actions $\textit{move}(i)$s. More precisely, given an history $h\in H$, strategy $\sigma_{\mathbf{p}}$ gives the probability with which the patroller will move to vertices at the next turn. Such a probability can be strictly positive only for vertices that are adjacent to that where the patroller is after history $h$. Notice that the patroller's strategy does not depend on the actions undertaken by the intruder. This is because the patroller cannot observe them. (Once the patroller has observed the intruder, i.e., when the patroller captures the intruder, the game concludes.) When $\sigma_{\mathbf{p}}$ is in pure strategies, i.e., when $\sigma_{\mathbf{p}}$ assigns all the probability to a single vertex for each possible history $h$, we say that the patrolling strategy is \emph{deterministic}. For instance, considering Figure~\ref{F:graforunningexample} a deterministic strategy could prescribe the patroller to follow the cycle $\langle 04, 05, 06, 11, 18, 17, 16, 10, 04\rangle$. Otherwise, we say that the patrolling strategy is \emph{non-deterministic}. For instance, consider again Figure~\ref{F:graforunningexample} with the patroller in vertex $01$ after a history $h$, a non-deterministic strategy could be: 
\[\sigma_{\mathbf{p}}(h)=\begin{cases}01& \textnormal{ with probability 0.25}\\ 02 & \textnormal{ with probability 0.25}\\ 06& \textnormal{ with probability 0.5}\end{cases}.\]
We define the \emph{intruder's strategy} as $\sigma_{\mathbf{i}}:H\rightarrow \Delta(V\cup\{wait\})$ where $\Delta(V\cup\{wait\})$ is a probability distribution over the vertices $V$ or, equivalently, over the corresponding actions $\textit{enter}(i)$ and the action $\textit{wait}$.

\subsection{Leader-Follower Solution Concept}
\label{subsez:stackelbger}

The intruder's ability to observe the patroller's strategy and act on the basis of such observation ``naturally'' induces one to analyze the game from a leader-follower stance, where the patroller is the leader and the intruder is the follower. The appropriate solution concept for leader-follower games is the \emph{leader-follower} (also said \emph{Stackelberg}) equilibrium~\cite{commitmentbased}. Its peculiarity is that the leader commits to a strategy and the intruder acts  as a best responder given such commitment. Rigorously speaking, the follower is not just a best responder: in order to have an equilibrium, if it is indifferent between some actions, it should choose the one that maximizes the patroller's expected utility. In~\cite{commitmentbased}, the authors show that in any two-player strategic-form game the leader never gets worse by committing to a leader-follower equilibrium strategy than by playing a Nash equilibrium strategy. Therefore, the leader will always commit to a leader-follower equilibrium strategy. However, to the best of our knowledge, there is not any similar result for the game we are dealing with (two-player extensive-form game with imperfect information and infinite horizon). In what follows  we extend the result presented in~\cite{commitmentbased}  to our game.

First, we consider the patroller's strategy in absence of any commitment, later we will show that the patroller never gets worse when it commits to a leader-follower equilibrium strategy. The appropriate solution concept for an extensive-form game with imperfect information is the \emph{sequential equilibrium}~\cite{krepswilson1982}, which is a refinement of Nash equilibrium. More precisely, a sequential equilibrium is a pair $(\sigma, \mu)$ where $\sigma$ is the agents' strategy profile and $\mu$ is a \emph{system of beliefs} (it prescribes how agents update their beliefs during the game). In a sequential equilibrium, the strategies are guaranteed to be rational (\emph{sequential rationality}) and the beliefs to be consistent with the agents' optimal strategies (\emph{Kreps and Wilson's consistency}). The presence of an infinite horizon complicates the study of the game. Considering our model, the patroller's strategy $\sigma_{\mathbf{p}}(h)$ is in principle infinite, $h$ being in principle infinitely long. With an infinite horizon, classic game theory studies a game by introducing symmetries, e.g., an agent will repeat a given strategy every $\bar{k}$ turns. (A classical example is the Rubinstein's alternating-offers protocol~\cite{rubinsteinECON1982}, where a buyer and a seller can negotiate without any deadline.) Introducing symmetries in our model amounts to fix the length $l$ of the histories in $H$. That is, in the patroller's strategies, the next action is selected on the basis of the last $l$ actions, with $l$ finite and constant during all the game. For instance, when $l=0$, actions in the patroller's strategy do not depend on any previously taken action. Namely, the probability to visit a vertex does not depend on the (adjacent) vertex where the patroller currently is. Hence, when $l=0$, the patroller performs well only in environments that are fully connected. When $l=1$, the patroller chooses its next action on the basis of its last action and then its strategy is Markovian. In this case, the selection of the next vertex to visit depends only on the current vertex of the patroller. Reasonably, when increasing the value of $l$, the expected utility of the patroller never decreases, because the patroller considers more information to select its next action. Classic game theory shows that infinite horizon extensive-form games admit a maximum length, say $\overline{l}$, of the symmetries such that beyond $\overline{l}$ the expected utility does not increase anymore~\cite{fudenberg1991}. In our model, this means that when the patroller's strategy is defined on the last $l$ vertices with $l\geq \overline{l}$ the patroller's expected utility is the same it receives when $l=\overline{l}$. On the other hand, we expect that when increasing the value of~$l$, the computational complexity for finding a patrolling strategy increases. This is because the spaces of patroller's and intruder's strategies become larger. In particular, the number of possible pure strategies $\sigma_{\mathbf{p}}(h)$s and $\sigma_{\mathbf{i}}(h)$s is $O(n^l)$, where $n$ is the number of vertices.  In practical settings, the selection of a value for $l$ is a trade-off between expected utility and computational effort.

Given a value for $l$, our game can be essentially reduced to a strategic-form game, because the game repeats every $l$ turns. Therefore, we can consider a reduced game that is $l$-turn long and constrain agents' possible strategies to be indefinitely repeated in all $l$-turn games. In our case, the patroller's strategy $\sigma_{\mathbf{p}}$ can be represented by a collection of $\{\alpha_{h,i}\}$, where $\alpha_{h,i}$ is the probability to execute action \emph{move($i$)} given history $h$. The intruder's strategy $\sigma_{\mathbf{i}}$ can be conveniently represented by using the following macro-actions: $\textit{enter-when}(i,h)$ and $\textit{stay-out}$. Action $\textit{enter-when}(i,h)$ corresponds to make $\textit{wait}$ until the patroller has followed history $h$ and then to make $\textit{enter}(i)$; $\textit{stay-out}$ corresponds to make $\textit{wait}$ forever.

We now show that when the leader (in our case the patroller) commits to a leader-follower equilibrium it obtains an expected utility that is not worse than the expected utility it would obtain from a sequential equilibrium. 
\begin{thm}
Given the game described above with a fixed $l$, the leader never gets worse when committing to a leader-follower equilibrium strategy with respect to a sequential equilibrium strategy.
\end{thm}
\emph{Proof.} The finite horizon game we obtain after having introduced symmetries can be easily translated into a strategic-form game, as prescribed by the classic normal-form of an extensive-form game~\cite{fudenberg1991}. If the leader does not commit to any strategy, it receives the expected utility prescribed by a sequential equilibrium of the game. This equilibrium is a specific Nash equilibrium of the strategic-form game. By von Stengel and Zamir~\cite{commitmentbased}, in any two-player strategic-form game, the worst leader-follower equilibrium is not worse than the best Nash equilibrium for the leader, and therefore, in our case, the worst leader-follower equilibrium is not worse than any sequential equilibrium for the leader (patroller). The thesis of the theorem follows.\hfill$\Box$

According to the above theorem, in our patrolling setting the patroller (leader) will commit to the leader-follower equilibrium strategy, which is the optimal patrolling strategy. 

\subsection{Summary of the Main Results}
\label{subsection:summaryofresults}

The main contribution of this paper is an algorithm to compute the optimal patrolling strategy (i.e., the leader-follower equilibrium strategy) for the setting we consider. We present our algorithm in Sections~\ref{S:deterministic} and~\ref{section:nondeterministic}. For the sake of presentation, we discuss our algorithm under the assumption that the setting is such that the optimal patrolling strategy gives the patroller a non-null probability of visiting all the targets. Otherwise, at least a target would be never visited (when following the optimal patrolling strategy) and the intruder will always have success in entering that target. In these cases, multiple robots should be employed. We will see in Section~\ref{S:extensions} how our approach can be used to capture some multirobot settings.

One of the most important features we considered in developing the proposed algorithm is computational efficiency. In order to achieve good performance in the process of finding a leader-follower equilibrium, and thus an optimal patrolling strategy, we divided this process in two steps. \textcolor{black}{In the first one, the algorithm searches for a deterministic patrolling strategy such that the intruder's best response is to make $\textit{stay-out}$. If such a strategy exists, then it is a leader-follower equilibrium strategy because it provides the patroller with the largest utility. Instead, if such a strategy does not exist, then the algorithm proceeds with the second step where it searches for an equilibrium characterized by a mixed, or non-deterministic, strategy.} The main reason behind this two-step process is that searching for a deterministic equilibrium strategy is computationally less expensive than searching for a non-deterministic one. Therefore, solving the first step separately leads to a significant improvement in computational performances every time a deterministic equilibrium strategy can be found. In the following we summarize the details of the two steps.

In Section~\ref{S:deterministic} the algorithm for computing a deterministic equilibrium patrolling strategy is presented. A deterministic equilibrium strategy is found as the solution of a feasibility problem, which is computed by resorting to constraint programming techniques. When solving this problem, the patrolling setting can be significantly simplified by reducing the graph to target vertices only. Moreover, another important advantage coming from this formulation is that it does not depend on the history length, namely, the value of $l=|h|$ can be arbitrarily large.

In Section~\ref{section:nondeterministic} the non-deterministic case is discussed. When the problem is to find a non-deterministic equilibrium strategy, the patrolling setting must be entirely considered, differently from the previous case. Moreover, the mathematical formulation of the problem strongly depends on the history length  $l$, yielding significant influences on the computational effort needed to compute its solution. The algorithm we propose works in three steps. First, in order to simplify the resolution process, all the agents' dominated actions, i.e., the actions that a rational agent would never play, are discarded. In the second step, the algorithm searches for an equilibrium where the intruder's strategy is \emph{stay-out} and, if such equilibrium is found, the corresponding non-deterministic strategy of the patroller is taken as the optimal patrolling strategy. Conversely, if the second step does not find an admissible solution, the algorithm proceeds with the last step where a leader-follower equilibrium is computed and the optimal patrolling strategy is found. The problem addressed in the second step is formulated as a bilinear feasibility mathematical programming problem while that addressed in the third step is formulated as a multi-bilinear optimization problem.


\section{Finding Deterministic Equilibrium Patrolling Strategies}
\label{S:deterministic}
In this section, we formally state the problem of finding a deterministic equilibrium patrolling strategy (Section~\ref{S:deterministic:problem}), we survey the main related works (Section~\ref{S:deterministic:sota}), we study the computational complexity of this problem (Section~\ref{S:deterministic:complexity}), we provide our solving algorithm (Section~\ref{S:deterministic:solving}), and we experimentally evaluate it (Section~\ref{S:deterministic:experiments}).

\subsection{Problem Formulation}
\label{S:deterministic:problem}
A \emph{deterministic patrolling equilibrium strategy} $\sigma_{\mathbf{p}}$ is represented as a sequence of vertices, or equivalently as a sequence of actions $move(i)$, such that, when adopted by the patroller, each target is left uncovered for a number of turns not larger than its penetration time and thus the optimal action of the intruder is $\textit{stay-out}$. Indeed, if the intruder attempts to enter a target, it will be captured by the patroller. (We recall that we are interested in studying situations wherein the equilibrium patrolling strategy covers all the targets.) For algorithmic reasons, we will consider deterministic equilibrium patrolling strategies that are \emph{cyclic}, namely that are composed of a finite sequence of vertices starting and ending with the same vertex, such that, when indefinitely repeated by the patroller, make entering any target not rational for the intruder.  As we shall show below, our representation of deterministic patrolling strategies does not fix any upper bound on the length $l$ of history $h$. 

At first, we show that the problem of searching for a deterministic equilibrium strategy in a graph $G$ is equivalent to search for a deterministic equilibrium strategy in a reduced graph $G'$. More precisely, we show that if $G'$ admits a deterministic equilibrium strategy $\sigma_{\mathbf{p}}'$, then we can always derive from $\sigma_{\mathbf{p}}'$  a strategy $\sigma_{\mathbf{p}}$ that is a  deterministic equilibrium strategy  for $G$, and, if $G'$ does not admit any  deterministic equilibrium strategy, then also $G$ does not. Reducing $G$ to $G'$, searching for a deterministic equilibrium strategy $\sigma_{\mathbf{p}}'$ for $G'$, and deriving $\sigma_{\mathbf{p}}$ from $\sigma_{\mathbf{p}}'$ will allow us to save a large amount of computational time with respect to searching directly for $\sigma_{\mathbf{p}}$ in $G$.  We start by discussing how we can reduce $G$ to $G'$.

\begin{figure}[h]
\centering
\begin{minipage}{5.5cm}
\centering \includegraphics[scale=0.6]{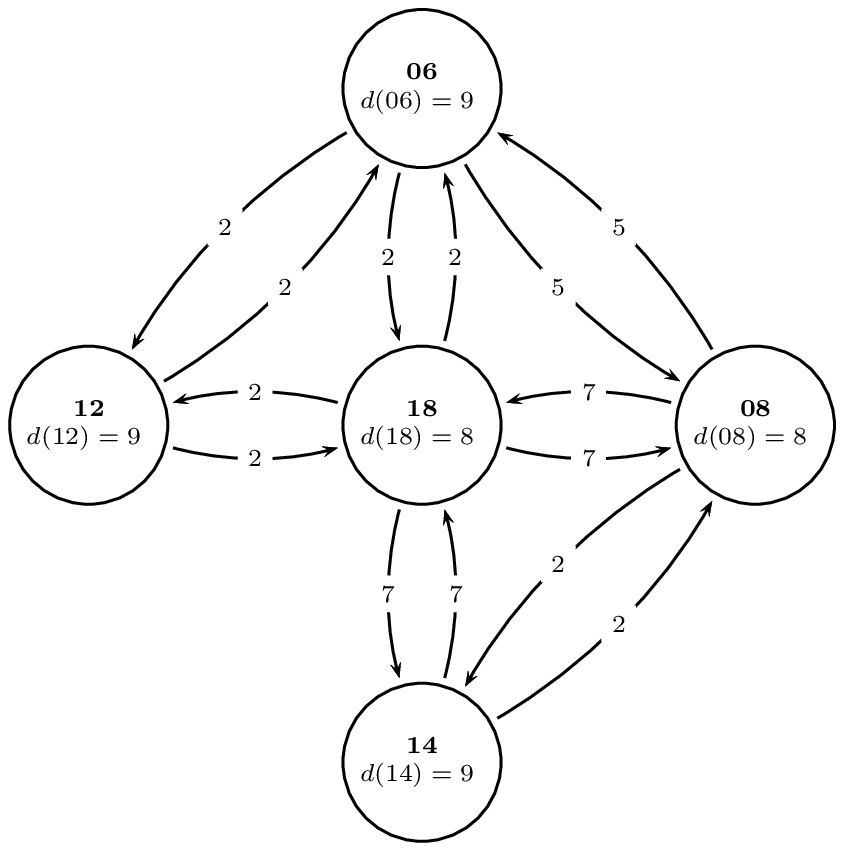}\\
\centering \scriptsize ({a})
\end{minipage}
\hspace{0.1cm}
\begin{minipage}{5.5cm}
\centering \includegraphics[scale=0.6]{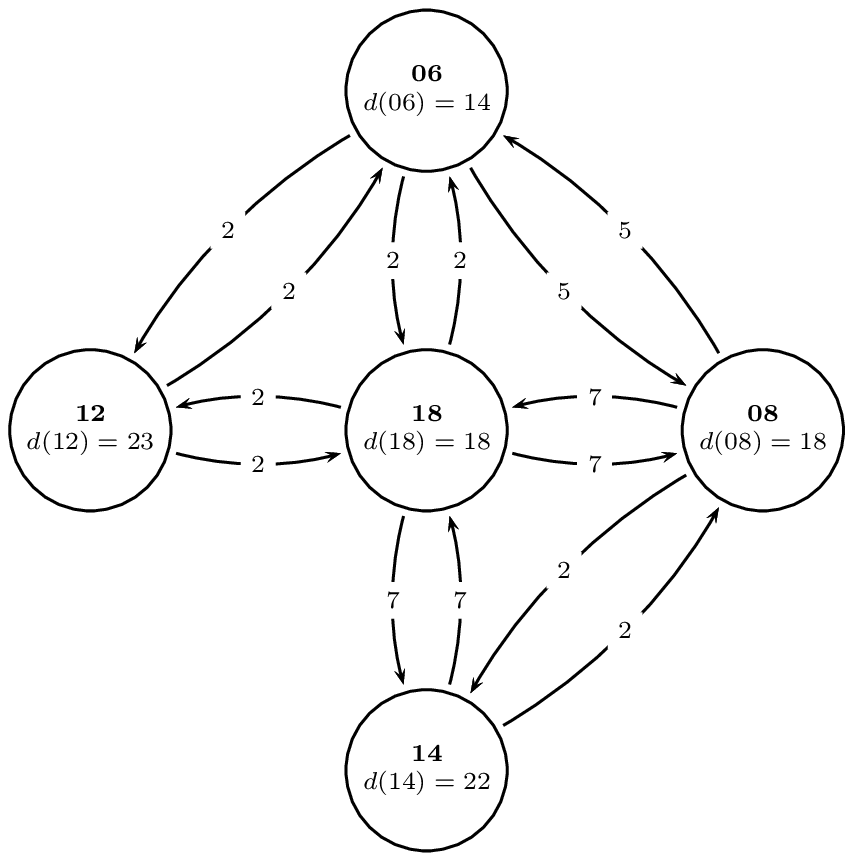}\\
\centering \scriptsize  ({b})
\end{minipage}
\caption{({a}) Reduced graph $G'$ corresponding to that of Figure~\ref{F:graforunningexample}. ({b}) The same graph as in ({a}), but with different penetration times.}
\label{F:graforidotto}
\end{figure}

The idea at the basis of reduction is that $G'$ is composed only of targets and the patroller will move between two targets along a shortest path connecting them. Formally, starting from $G=(V,A,T,v,d)$, we define $G'=(T,A',w,d)$, where targets $T$ are the vertices of $G'$; $A'$ is the set of arcs connecting the targets defined as a function $a':T\times T\rightarrow\{0,1\}$ and derived from set $A$ as follows: for every pair of targets $i,j\in T$ and $i \neq j$, $a'(i,j)=1$ if at least one of the shortest paths connecting $i$ to $j$ in $G$ does not pass through any other target, $a'(i,j)=0$ otherwise; $w$ is a weight function defined as $w:T\times T\rightarrow \mathbb{N}$ where $w(i,j)$ is the length of the shortest path between $i$ and $j$ in $G$ ($w(i,j)$ is defined only when $a'(i,j)=1$); $d$ is defined as in $G$. Notice that agents' values do not appear in $G'$. This is because the computation of a deterministic equilibrium strategy covering all the targets does not depend on their evaluations. The reduction algorithm develops in two steps.
\begin{enumerate}
\item The shortest paths connecting each pair of targets are computed by repeatedly applying Dijkstra's algorithm~\cite{cormen2001} to $G$. (Dijkstra's algorithm is applicable since no negative arc weight appears in $G$.) The asymptotical worst-case computational complexity is $O(t n^2)$, where $t$ is the number of targets in $T$ and $n$ is the number of vertices in $V$.
\item All the shortest paths computed in the previous step are analyzed and, for every path that does not contain targets (except from the source and the destination), the corresponding arc is added to $A'$ (avoiding duplications) and its weight $w$ is set equal to the length of the shortest path. This is accomplished by applying a linear search to each shortest path with a resulting computational complexity of $O(t^2 n)$.
\end{enumerate}
Consider the graph reported in Figure~\ref{F:graforunningexample}. The corresponding reduced graph $G'$ is reported in Figure~\ref{F:graforidotto}(a). $G'$ is composed of only 5 vertices. (We report another graph in Figure~\ref{F:graforidotto}(b) that differs from that in ({a}) in the penetration times; we shall use it as example in the following sections.) 

We now show that given a strategy $\sigma_{\mathbf{p}}'$ for $G'$ we can derive a strategy $\sigma_{\mathbf{p}}$ for $G$. Specifically, $\sigma_{\mathbf{p}}$ is built such that, for any pair of consecutive vertices $(i,j)$ in $\sigma_{\mathbf{p}}'$, there is a sequence of vertices $(i,z_1,\ldots,z_m,j)$ in $\sigma_{\mathbf{p}}$ that is a shortest path connecting $i$ to $j$. We can now state the following proposition.
\begin{prop}
If $G'$ admits a deterministic equilibrium strategy $\sigma_{\mathbf{p}}'$, then $\sigma_{\mathbf{p}}$ (derived as discussed above) is a deterministic equilibrium strategy for $G$, and, if $G'$ does not admit any  deterministic equilibrium strategy, then also $G$ does not.
\end{prop}
\emph{Proof.}
The proof of the first implication is trivial. Indeed, given any deterministic equilibrium strategy $\sigma_{\mathbf{p}}'$ defined on $G'$, the corresponding  $\sigma_{\mathbf{p}}$ is a deterministic equilibrium strategy in $G$, because the time (number of turns) between two successive visits to a target in  $\sigma_{\mathbf{p}}'$ is the same that the time between two successive visits to the same target in  $\sigma_{\mathbf{p}}$ (by definition of weights $w$ in $G'$). To demonstrate the second implication we proceed by contradiction assuming that $G$ admits a solution and that $G'$ does not. This means that the solution in $G$ prescribes the patroller to move between some targets along paths different from the shortest ones. However, if no solution that uses shortest paths can cover all the targets without leaving any one of them uncovered for more than its penetration time, then no other solution can do it, because, not using shortest paths, time between two successive visits to a target can only increase. Therefore, it follows by contradiction that if $G'$ does not admit any solution, then also $G$ does not.\hfill$\Box$

We are now in the position to formally define the problem of searching for a deterministic equilibrium strategy. For the sake of presentation, from here on we use $\sigma$ in the place of $\sigma_{\mathbf{p}}'$ and we talk of vertices of $G'$ and $\sigma$, instead of targets. Formally, we define a function $\sigma:\{1,2,\ldots,s\}\rightarrow T$, where $\sigma(j)$ is the $j$-th element of the sequence. The length of the sequence is $s$. The \emph{temporal length} of a sequence of visits is computed by summing the weights of covered arcs, i.e.,  by summing the times (in number of turns) for covering arcs, $\sum_{j=1}^{s-1}w\left(\sigma(j),\sigma(j+1)\right)$. The time interval between two visits of a vertex is calculated similarly, summing the weights of the arcs covered between the two visits.

A \emph{solution} of our problem (a deterministic equilibrium patrolling strategy) is a sequence $\sigma$ such that the following properties are satisfied:
\begin{enumerate}
\item $\sigma$ is cyclical, i.e., the first vertex coincides with the last one, namely, $\sigma(1) = \sigma(s)$;
\item every vertex in $T$ is visited at least once, i.e., there are no uncovered vertices;
\item when indefinitely repeating the cycle, for any $i \in T$, the time interval between two successive visits of $i$ is never larger than $d(i)$.
\end{enumerate}
It is clear that, when the patroller follows repeatedly such a sequence $\sigma$ as its patrolling strategy, no intrusion can occur.

Let denote by $O_i(j)$ the position in $\sigma$ of the $j$-th occurrence of $i$ and by $o_i$ the total number of $i$'s occurrences in $\sigma$. For instance, consider Figure~\ref{F:graforidotto}(a): given $\sigma=\langle14,08,18,08,14\rangle$, $O_{08}(1)=2$ and $O_{08}(2)=4$, while $o_{08}=2$ and $o_{06}=0$. Notice that, given a sequence $\sigma$, quantities $O_i(j)$ and $o_i$ can be easily calculated.

Given such definitions we can formally re-state the problem in a mathematical programming fashion. We aim at finding a sequence $\sigma$ of $s$ visits to vertices of $G'$ (targets) such that the following constraints are satisfied:

\begin{scriptsize}
\begin{eqnarray}
\sigma(1)=\sigma(s)& \label{cond:iniziougualefine}\\
o_i\geq1 & \forall i\in T \label{cond:tuttidentro}\\
a'\left(\sigma(j-1),\sigma(j)\right)=1 & \forall j\in\{2,3,\ldots,s\} \label{cond:raggiungibile}\\
\sum_{j=O_i(k)}^{O_i(k+1)-1}w\left(\sigma(j),\sigma(j+1)\right)\leq d(i) & ~~~~\forall i\in T,\forall k\in\{1,2,\ldots,o_i-1\}~~~~\label{cond:intermezzi}\\
\sum_{j=1}^{O_i(1)-1}w\left(\sigma(j),\sigma(j+1)\right)+\sum_{j=O_i(o_i)}^{s-1}w\left(\sigma(j),\sigma(j+1)\right)\leq d(i) & \forall i\in T \label{cond:chiusuraciclo}
\end{eqnarray}
\end{scriptsize}

\noindent Constraint~(\ref{cond:iniziougualefine}) states that $\sigma$ is a cycle, i.e., the first and last vertices of $\sigma$ coincide; constraints~(\ref{cond:tuttidentro}) state that every vertex is visited at least once in $\sigma$; constraints~(\ref{cond:raggiungibile}) state that for every pair of consecutively visited vertices, say $\sigma(j-1)$ and $\sigma(j)$, it is $a'(\sigma(j-1),\sigma(j))=1$, i.e., vertex $\sigma(j)$ can be directly reached from vertex $\sigma(j-1)$ in $G'$; constraints~(\ref{cond:intermezzi}) state that, for every vertex $i$, the temporal interval between two successive visits of $i$ in $\sigma$ is not larger than $d(i)$; similarly, constraints~(\ref{cond:chiusuraciclo}) state that for every vertex $i$ the temporal interval between the last and first visits of $i$ is not larger than $d(i)$, i.e., the deadline of $i$ must be respected also along the cycle closure. Hence, our goal is to find a sequence of vertices $\sigma(1)$, $\sigma(2)$, \ldots, $\sigma(s)$ such that the above constraints are satisfied. Notice that also the length $s$ of the sequence must be found as part of the solution. In Section~\ref{S:deterministic:solving}, we present a method for finding a sequence $\sigma$ that satisfies constraints (\ref{cond:iniziougualefine})-(\ref{cond:chiusuraciclo}).


If we consider, for instance, the problem described by the graph of Figure~\ref{F:graforidotto}(a), it is easy to show that no sequence of visits can satisfy all the constraints described above. To see this, it is enough to observe that the shortest cycle covering only vertices $06$ and $08$, i.e., $\langle06,08,06\rangle$, has a temporal length larger than the penetration times of both the involved vertices, so there is no way to cover these vertices (and others) within their penetration times. As we shall show below, the problem described by the graph of Figure~\ref{F:graforidotto}(b) admits a deterministic equilibrium strategy.

\subsection{Related Works}
\label{S:deterministic:sota}
To the best of our knowledge, the strategic patrolling literature does not present significant works for finding a deterministic equilibrium strategy. In particular, it does not take into account the penetration times of the targets, see, e.g.,~\cite{yannIAT2004,gladECAI2008}. A larger amount of works related to our problem can be found in the operational research literature, since the problem is ultimately that of finding a particular sequence of vertices in the graph $G'$. The main of these works are extensions and refinements of the Traveling Salesman Problem (TSP). The first extension of the TSP we consider relates to settings with temporal constraints and is called \emph{deadline-TSP}~\cite{tsitsiklis1992}. In this problem vertices have deadlines over their first visit and some time is spent traversing arcs. Rewards are collected when a vertex is visited before its deadline, while penalties are assigned when a vertex is either visited after its deadline or is not visited at all. The objective is to find a tour that maximizes the reward, visiting as many vertices as possible. Vertices can be visited more than once in the tour, but the reward/penalty is received only at the first visit. 
A more general variant is the \emph{Vehicle Routing Problem with Time Windows}~\cite{kolen1987} where deadlines are replaced with time windows, during which visits of vertices must occur. This problem has been studied also by employing constraint programming techniques~\cite{CSPVRPTW}. Cyclical sequences of visits are addressed in the \emph{period routing problem}~\cite{christofides1984} where vehicle routes are constructed to run for a finite period of time in which every vertex has to be visited according to a given frequency. Frequencies can be given also as lower bounds, considering the real frequencies of visits as decision variables of the problem~\cite{francis2006}. In the \emph{cyclic inventory routing problem}~\cite{raa2009} vertices represent customers with a given demand rate and storage capacity. The objective is to find a tour such that a distributor can repeatedly restock customers under some constraints over visiting frequencies.

Two are the main issues that distinguish our problem from those described above. The first one is that our problem is defined according to relative deadlines (calculated with respect to two consecutive visits to the same vertex) and the absolute deadlines (calculated with respect to the beginning of the sequence) depend on the solution itself, as expressed by constraints~(\ref{cond:intermezzi})-(\ref{cond:chiusuraciclo}).  The extension of the above works, mostly considering absolute deadlines, to our problem introduces highly non-linear constraints and does not seem to be straightforward. The second issue is that in our problem we aim at finding only \emph{a} solution, and not the optimal solution according to some metric. Thus, we are solving a feasibility problem and not an optimization problem.

\subsection{NP-Completeness}
\label{S:deterministic:complexity}
In this section we discuss the computational complexity of finding a deterministic equilibrium strategy for our patrolling setting, denoted as the DET-STRAT problem.

\begin{thm} The DET-STRAT problem is NP-complete.\end{thm}
\emph{Proof}. We prove the NP-completeness by reducing the Directed Hamiltonian Circuit problem (DHC)~\cite{compcomplexity} to the DET-STRAT problem. DHC is the problem of determining if an Hamiltonian path, i.e., a path that visits each vertex exactly once, exists in a given directed graph. This is a well-known NP-complete problem. (For the sake of simplicity, we consider a graph $G=(V,A,T,v,d)$. A similar, but more complicated, proof can be produced considering a graph $G'=(T,A,w,d)$.) Let us consider a generic instance of the DHC problem given by a directed graph $G_h=(V_h,A_h)$ where $V_h$ is the set of vertices and $A_h$ is the set of arcs. In order to prove that DHC can be reduced to the DET-STRAT problem, we show that for every instance $G_h$ of the DHC problem an instance $G_s$ of the DET-STRAT problem can be built in polynomial time and that by solving the DET-STRAT problem on $G_s$ we obtain also a solution for the DHC problem on $G_h$. An instance $G_s = (V_s,A_s,T_s,v_s,d_s)$ can be easily constructed from $G_h$ in the following way: $V_s=T_s=V_h$, $A_s=A_h$, for every $v \in V_s$ we impose $d(v) = |V_h|$ and the functions $v_s$ can be arbitrarily defined. It is straightforward to see that a solution of $G_s$, if it exists, is an Hamiltonian cycle. Indeed, since the relative deadline of every target is equal to the number of targets, a deterministic equilibrium strategy should visit each target exactly once, otherwise at least one relative deadline would be violated (recall that covering one arc in $G_{s}$ is assumed to require one time unit). Therefore, computing the solution for $G_s$ provides by construction a solution for $G_h$ or, in other words, the DHC problem can be reduced to the DET-STRAT problem, proving its NP-completeness (it is trivially polynomial to verify that a given sequence of vertices is a solution of the DET-STRAT problem).\hfill$\Box$

\subsection{Solving Algorithm}
\label{S:deterministic:solving}
In this section, we present our basic solving algorithm (Section~\ref{S:deterministic:solving:basic}), we report an example (Section~\ref{S:deterministic:solving:example}), we theoretically analyze some properties of the algorithm (Section~\ref{S:deterministic:solving:analysis}), and we show how to improve its efficiency (Section~\ref{S:deterministic:solving:improving}).

\subsubsection{Basic Algorithm}
\label{S:deterministic:solving:basic}
We formulate the problem of finding a deterministic equilibrium patrolling strategy described in the Section~\ref{S:deterministic:problem} as a Constraint Satisfaction Problem (CSP) \cite{russel2003}. Each $\sigma(j)$ is considered as a variable with domain $F_j\subseteq T$. The constraints over the values of the variables are~(\ref{cond:iniziougualefine})-(\ref{cond:chiusuraciclo}). A solution is an assignment of values to all the variables such that all those constraints are satisfied. When the problem is put in this form, it resembles some problems of CSP-based scheduling. When compared with the large literature in AI that studies scheduling problems, cyclic scheduling presents comparatively limited results (e.g.,~\cite{IJCAI1999}). Differently from the problems studied in the literature, in our problem the number $s$ of variables $\sigma(j)$s which compose the solution is not known in advance and must be computed as part of the solution. This is because a vertex can appear more times in $\sigma$. As a consequence, we cannot resort to constrain programming tools, e.g., ILOG~CP~\cite{ilogcp}, and we need to develop an \emph{ad hoc} algorithm.

The algorithm we propose for finding a solution basically searches the state space with backtracking. Forward checking~\cite{russel2003} is used in the attempt to reduce the branching of the search tree. We report our algorithm in Algorithms~\ref{A:init},~\ref{A:recursivecall}, and~\ref{A:forwardchecking}.

Algorithm~\ref{A:init} simply assigns $\sigma(1)$ a vertex $i \in T$. Notice that if a solution exists, it can be found independently of the first vertex appearing in $\sigma$. Since the solution $\sigma$ is a cycle that visits all vertices, every vertex can be chosen as the initial one. Hence, the choice of $i$ in Algorithm~\ref{A:init} does not affect the possibility of finding a solution.

\begin{algorithm2e}[h]
\scriptsize \caption{\textsc{find\_solution($T,A',w,d$)}} \label{A:init}
\dontprintsemicolon \BlankLine
select a vertex $i$ in $T$

assign $\sigma(1)\leftarrow i$

call \textsc{recursive\_call}($T,A',w,d,\sigma,2$)
\end{algorithm2e}

Algorithm~\ref{A:recursivecall} assigns $\sigma(j)$ a vertex from domain $F_j\subseteq T$, which contains available values for $\sigma(j)$ that are returned by the forward checking algorithm (Algorithm~\ref{A:forwardchecking}). If $F_j$ is empty or no vertex in $F_j$ can be successfully assigned to $\sigma(j)$, then Algorithm~\ref{A:recursivecall} returns failure and a backtracking is performed.

\begin{algorithm2e}[h]
\scriptsize \caption{\textsc{recursive\_call($T,A',w,d,\sigma,j$)}} \label{A:recursivecall}
\dontprintsemicolon \BlankLine
\If{$\sigma(1)=\sigma(j-1)$ and constraints (\ref{cond:tuttidentro}) hold}
{\If{constraints (\ref{cond:chiusuraciclo}) hold}{return $\sigma$} \Else{return \textsc{failure}}}
\Else{assign $F_j\leftarrow \textsc{forward\_checking}(T,A',w,d,\sigma,j)$

\For{all the $i$ in $F_j$}
{assign $\sigma(j)\leftarrow i$

assign $\sigma'\leftarrow \textsc{recursive\_call}(T,A',w,d,\sigma,j+1)$

\If{$\sigma'$ is not \textsc{failure}} {return $\sigma'$}
}

return \textsc{failure}
}
\end{algorithm2e}

Algorithm~\ref{A:forwardchecking} restricts $F_j$ to the vertices directly reachable from the last assigned vertex $\sigma(j-1)$ such that their visits do not violate constraints~(\ref{cond:intermezzi})-(\ref{cond:chiusuraciclo}). Notice that checking constraints~(\ref{cond:intermezzi})-(\ref{cond:chiusuraciclo}) requires knowing the weights (temporal costs) related to the arcs between the vertices that could be assigned subsequently, i.e., between the variables $\sigma(k)$ with $k>j$. For example, consider the graph of Figure~\ref{F:graforidotto}(b) and suppose that the partial solution currently constructed by the algorithm is $\sigma=\langle 14 \rangle$. In this situation, we cannot check the validity of constraints~(\ref{cond:intermezzi})-(\ref{cond:chiusuraciclo}) since we have no information about times to cover the arcs between the vertices that will complete the solution. In order to cope with this, we estimate the unknown temporal costs by employing an admissible heuristic (i.e., a non-strict underestimate) based on the minimum cost between two vertices. The heuristic being admissible, no feasible solution is discarded. We denote the heuristic value by $\overline{w}$, e.g., $\overline{w}(i,\sigma(1))$ denotes the weight of the shortest path between $i$ and $\sigma(1)$. We assume $\overline{w}(i,i) = 0$ for any $i$.

Given a partial solution $\sigma$ from $1$ to $j-1$, the forward checking algorithm considers all the vertices directly reachable from $\sigma(j-1)$ and keeps those that do not violate the relaxed constraints~(\ref{cond:intermezzi})-(\ref{cond:chiusuraciclo}) computed with heuristic values. It considers  a vertex $i$ directly reachable from $\sigma(j-1)$ and assumes that $\sigma(j)=i$. Step~5 of Algorithm~\ref{A:forwardchecking} checks relaxed constraints~(\ref{cond:chiusuraciclo}) with respect to $i$, assuming that the weight along the cycle closure from $\sigma(j)=i$ to $\sigma(1)$ is minimum. In the above example, with $\sigma(1)=14$, the vertices directly reachable from $\sigma(1)$ are $08$ and $18$. The algorithm considers $\sigma(2) = 08$. By Step~5, we have $w(\sigma(1),08)+\overline{w}(08,\sigma(1))=4\leq d(08)=18$ and then Step~5 is satisfied. It can be easily observed that such condition holds also at the next iteration of the cycle, when $\sigma(2)=18$. Step~8 of Algorithm~\ref{A:forwardchecking} checks relaxed constraints~(\ref{cond:chiusuraciclo}) with respect to all the vertices $k\neq i$, assuming that both the weight to reach $k$ from $\sigma(j)=i$ and the weight along the cycle closure from $k$ to $\sigma(1)$ are minimum. Consider again the above example. It can be easily observed that when $\sigma(2) = 08$ such conditions hold for all~$k$. Instead, at the next iteration, when $\sigma(2)=18$ and $k=06$ we have $w(\sigma(1),18)+\overline{w}(18,06)+\overline{w}(06,\sigma(1))=16 > d(06)=14$. The relaxed constraint is violated and vertex $18$ will be not inserted in $F_{j}$. Similarly, Step~6 checks relaxed constraints~(\ref{cond:intermezzi}) with respect to $i$ and Step~9 checks relaxed constraints~(\ref{cond:intermezzi}) with respect to any $k$ assuming that  the weight to reach $k$ from $\sigma(j)=i$ is minimum. In the above example, starting from $\sigma = \langle 14 \rangle$, the relaxed constraints are satisfied only when $i=08$ and therefore $F_j=\{08\}$. Finally, we notice that Steps~5 and~8 are checked only when $o_i=0$ and $o_k=0$, respectively, since it can be easily proved that when $o_i>0$ and $o_k>0$ these conditions always hold.

\begin{algorithm2e}[h]
\scriptsize \caption{\textsc{forward\_checking($T,A',w,d,\sigma,j$)}} \label{A:forwardchecking}
\dontprintsemicolon \BlankLine
assign $F_j\leftarrow\emptyset$

assign $s\leftarrow j-1$

\For{all members $i$ in $T$ such that $a'(\sigma(s),i)=1$}
{
\If{conditions\\
$\Big(o_i=0 \wedge \sum_{l=1}^{s-1}w(\sigma(l),\sigma(l+1))+w(\sigma(s),i)+\overline{w}(i,\sigma(1))\leq d(i)$  or \\
$o_i>0 \wedge\sum_{l=O_i(o_i)}^{s-1}w(\sigma(l),\sigma(l+1))+w(\sigma(s),i)\leq d(i)\Big)$ and, \\
for all $k\neq i$,\\
$\Big(o_k=0\wedge\sum_{l=1}^{s-1}w(\sigma(l),\sigma(l+1))+w(\sigma(s),i)+\overline{w}(i,k)+\overline{w}(k,\sigma(1))\leq d(k)$ or\\
$o_k>0\wedge\sum_{l=O_k(o_k)}^{s-1}w(\sigma(l),\sigma(l+1))+w(\sigma(s),i)+\overline{w}(i,k)\leq d(k) \Big)$\\
hold}
{add $i$ to $F_j$}
}

return $F_j$
\end{algorithm2e}

\subsubsection{Example}
\label{S:deterministic:solving:example}

\begin{figure}[h]
\centering
\includegraphics[scale=0.85]{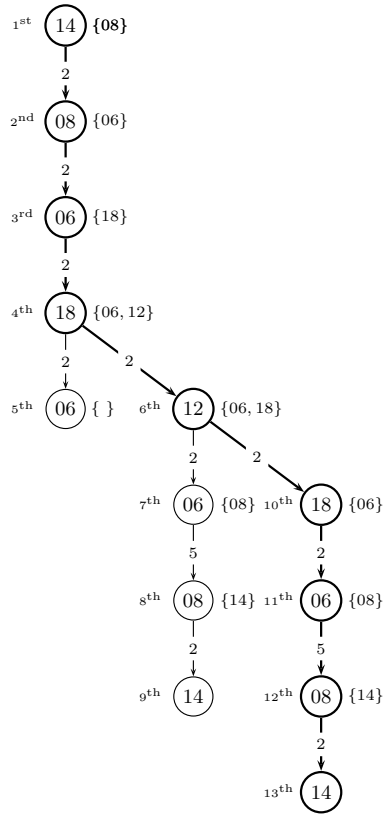} 
\caption{Search tree for the example of Figure~\ref{F:graforidotto}(b); bold nodes and arrows denote the obtained solution; $F_j$s are reported besides nodes $\sigma(j-1)$; $x^{th}$ denotes the order in which the tree's nodes are analyzed.}
\label{F:albero}
\end{figure}

We apply our algorithm to the example of Figure~\ref{F:graforidotto}(b). We use a random selection in Step~1 of Algorithm~\ref{A:init} (to choose the first visited vertex of the sequence) and in Step~7 of Algorithm~\ref{A:recursivecall} (to choose the elements of $F_j$ as part of the current candidate solution). We report part of the execution trace (Figure~\ref{F:albero} depicts the complete search tree):
\begin{enumerate}
\item[(a)] the algorithm assigns $\sigma(1)=14$;
\item[(b)] the domain $F_2$ (depicted in the figure between curly brackets beside vertex $\sigma(1)=14$) is produced as follows (recall the discussion of the previous section):
\begin{itemize}
\item vertex $08$ is added to $F_2$, since all the conditions in Algorithm~\ref{A:forwardchecking} with $i=08$ are satisfied;
\item vertex $18$ is not added to $F_2$, since the condition in Step~8 of Algorithm~\ref{A:forwardchecking} with $k=06$ is not satisfied, formally, $w(14,18)+\overline{w}(18,06)+\overline{w}(06,14)>d(06)$;
\item no other vertex is added to $F_2$, since no other vertex is directly reachable from $14$;
\end{itemize}
\item[(c)] the algorithm assigns $\sigma(2)=08$;
\item[(d)] the domain $F_3$ is produced similarly as above, yielding to $F_3=\{06\}$;
\item[(e)] the algorithm assigns $\sigma(3)=06$ and continues.
\end{enumerate}
Some issues are worth noting. In the $9^{th}$ node of the search tree, a sequence $\sigma$ with $\sigma(1)=\sigma(s)$ and including all the vertices was found. However, this sequence does not satisfy constraints~(\ref{cond:chiusuraciclo}). If the search is not stopped and backtracked at the the $9^{th}$ node (in Step~5 of Algorithm~\ref{A:recursivecall}), the algorithm would never terminate. Indeed, the subtrees that would follow this vertex would be the infinite repetition of part of the tree already built; in particular, of the solution in bold of Figure~\ref{F:albero}. Finally, in the $5^{th}$ node, no possible successor is allowed by the forward checking, and therefore the algorithm backtracks.  

\subsubsection{Some Properties of the Algorithm}
\label{S:deterministic:solving:analysis}

In this section, we present some properties of the proposed algorithm. At first, we prove its soundness and completeness.
\begin{thm} The algorithm of Section~\ref{S:deterministic:solving:basic} is sound and complete.\end{thm}
\emph{Proof}. We initially prove the soundness of the algorithm. We need to prove that all the solutions it produces satisfy constraints~(\ref{cond:iniziougualefine})-(\ref{cond:chiusuraciclo}). Constraints~(\ref{cond:iniziougualefine}), (\ref{cond:tuttidentro}), and~(\ref{cond:chiusuraciclo}) are satisfied by Algorithm~\ref{A:recursivecall}. If at least one of them does not hold, no solution is produced. The satisfaction of constraints~(\ref{cond:raggiungibile}) is assured by Algorithm~\ref{A:forwardchecking} in Step~3, while the satisfaction of constraints~(\ref{cond:intermezzi}) is assured by Algorithm~\ref{A:forwardchecking} in Steps~6 and~9.

In order to prove completeness we need to show that the algorithm produces a solution whenever at least one exists. In the algorithm there are only two points in which a candidate solution is discarded. The first one is the forward checking in Algorithm~\ref{A:forwardchecking}. Indeed, it iteratively applies constraints~(\ref{cond:intermezzi})-(\ref{cond:chiusuraciclo}) to a partial sequence $\sigma$ exploiting a heuristic over the future weights (i.e., the time spent to visit the successive vertices). Since the employed heuristic is admissible, no feasible candidate solution can be discarded. The second point is the stopping criterion in Algorithm~\ref{A:recursivecall}: when all the vertices occur in $\sigma$ (at least once) and the first and the last vertex in $\sigma$ are equal, no further successor is considered and the search is stopped. If $\sigma$ satisfies all the constraints, then $\sigma$ is a solution, otherwise backtracking is performed. We show that, if a solution can be found without stopping the search at this point, then a solution can be found also by stopping the search and backtracking (the \emph{vice versa} does not hold). This issue is of paramount importance since it assures that the algorithm terminates (recall the example of the previous section in which, without this stopping criterion, the search could not terminate). Consider a $\sigma$ such that $\sigma(1)=\sigma(s)$ and including all the vertices in $T$. The search subtree following $\sigma(s)$ and produced by the proposed algorithm is (non-strictly) contained in the search tree following from $\sigma(1)$. This is because the constraints considered by the forward checking from $\sigma(s)$ on are (non-strictly) harder than the corresponding ones from $\sigma(1)$ to $\sigma(s)$. The increased hardness is due to the activation of constraints~(\ref{cond:intermezzi}) that are needed given that at least one occurrence of each vertex is in $\sigma$. Thus, if a solution can be found by searching from $\sigma(s)$, then a shorter solution can be found by stopping the search at $\sigma(s)$ and backtracking. This concludes the proof of completeness.\hfill$\Box$

\begin{figure}[h]
\centering
\includegraphics{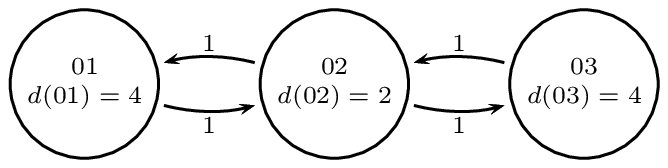}
\caption{A situation in which the temporally shortest solution is as long as the upper bound of Theorem~\ref{T:bound}.}
\label{F:worst}
\end{figure}

Now we derive an upper bound over the temporal length of $\sigma$.
\begin{thm}
If the problem defined in Section~\ref{S:deterministic:problem} admits a solution, then there exists at least a solution $\sigma$ with temporal length no longer than $\max_{t \in T}\{d(t)\}$.\label{T:bound}
\end{thm}
\emph{Proof.} In order to prove the theorem it is sufficient to prove that, if a problem is solvable, then there exists a solution $\sigma$ in which there is at least a vertex that only appears once, excluding $\sigma(s)$. Indeed, if this statement holds then the maximum temporal length of $\sigma$ is bounded by $d(i)$ where $i$ is the vertex that appears only one time in $\sigma$. It easily follows that, in the worst case, the maximum temporal length of $\sigma$ is $\max_{t \in T}\{d(t)\}$. Figure~\ref{F:worst} shows a situation in which the temporal length of the unique solution is exactly $\max_{t \in T}\{d(t)\}$.

We now prove that, if the problem is solvable, then there is a solution in which at least a vertex appears only once. To prove this, we consider a solution $\sigma$ wherein $\sigma(1)$ is the vertex with the minimum relative deadline, i.e,  $\sigma(1)={\arg\min}_{t \in T}\{d(t)\}$. (Notice that, according to the discussion of Algorithm~\ref{A:init}, this assignment does not preclude finding a solution.) We call $k$ the minimum integer such that all the vertices appear in the subsequence $\sigma(1)-\sigma(k)$. We show that, if the problem is solvable, then it is not necessary that vertex $v = \sigma(k)$ appears again after $k$. A visit to $v$ after $k$ would be observed if either it is necessary to pass through $v$ to reach $\sigma(1)$ or it is necessary to re-visit $v$, due to its relative deadline, before $\sigma(1)$. However, since all the vertices but $v =\sigma(k)$ are visited before $k$, all the vertices but $v$ can be visited without necessarily visiting $v$. Furthermore, the deadline of $\sigma(1)$ is by hypothesis harder than $\sigma(k)$'s one and then the occurrence of $v = \sigma(k)$ after $k$ is not necessary. Therefore, vertex $\sigma(k)$ occurs only one time.\hfill$\Box$

The above theorem provides an upper bound on the temporal length of~$\sigma$. Notice that there are cases for which this bound is exact; namely, there are instances of the problem for which the temporally shortest solution has length exactly $\max_{t \in T}\{d(t)\}$ (as in the graph of Figure~\ref{F:worst}). In other cases, the bound is not exact (as in Figure~\ref{F:worst} with $d(01) = 5$). In any case, the upper bound can be exploited to limit the depth of the search tree preserving the algorithm's completeness. Indeed, if $\sum_{l=1}^{s-1}w(\sigma(l),\sigma(l+1))+\overline{w}(\sigma(s),\sigma(1)) > \max_{t \in T}\{d(t)\}$, then the search can be safely stopped and backtracked.

Finally, we focus on linear settings, namely on situations in which a deterministic equilibrium strategy has to be found for a patroller acting in corridors (see, e.g., Figure~\ref{F:worst}) and rings. These settings appear commonly in real-world applications and their study can lead to the definition of simple heuristics that are very effective also for non-linear settings.
\begin{prop}
If a problem defined on a linear graph admits a solution, then the linear sequence of the vertices is a solution.\label{P:linear}
\end{prop}
\emph{Proof sketch.}  Consider a setting like that in Figure~\ref{F:worst}, with any functions $w$ and $d$. Suppose $\sigma(1)=01$ and consequently $\sigma(2)=02$. If there is no feasible $\sigma$ with $\sigma(3)=03$, then the problem is not feasible. Indeed, if $\sigma(3)=03$ does not satisfy the constraints over the deadline of $01$ along the cycle closure, then there is not any $k$ such that $\sigma(k)=03$ satisfies such constraints. The same argument can be applied to any other vertex and to any linear graph.\hfill$\Box$

The above proposition suggests a simple method to check the feasibility of a problem defined on a linear graph. If the linear sequence is not feasible, then the problem is unfeasible, otherwise it constitutes a solution. The length of such solution grows linearly in the number of vertices. Notice that the problem can admit more solutions, and the length of some of them can be larger than that of the linear solution. For example, in Figure~\ref{F:worst}, with $d(03)$ arbitrarily large, a solution could be $\sigma=\langle 01, 02, 01,02, \ldots, 03, 02, 01\rangle$.

Let us now consider our algorithm in linear settings. At first, we note that the problem is feasible if and only the application of the forward checking to the two extremes of the graph returns non-null domains. This is because, the conditions considered in the forward checking correspond exactly to the feasibility of the linear sequence and therefore, if the problem is unfeasible, the returned domain is empty. Furthermore, in order to guide efficiently the search it is sufficient that at each node of the search tree the successors are ordered from the minimum to the maximum $o_i$. In this case, the algorithm produces a search tree whose size is linear in the number of vertices.

\subsubsection{Improving Efficiency and Heuristics}
\label{S:deterministic:solving:improving}

In this section, we show how to reduce the number of constraints to be checked in the forward checking, we introduce a more sophisticated stopping criterion, and we propose some heuristics to select vertices.

Consider the conditions in Steps~5 and~8 of Algorithm~\ref{A:forwardchecking}.
Except for the first execution of Algorithm~\ref{A:forwardchecking} (i.e., when $j=2$), the satisfaction of the condition at Step~5 for a given $j$ is granted if the condition in Step~8 for $j-1$ is satisfied. Therefore, we can safely limit the algorithm to check the conditions at Step~5 exclusively when $j=2$. The same considerations hold also for the conditions in Steps~6 and~9. Therefore, we can safely limit the algorithm to check the conditions at Step~6 exclusively when $j=2$.

We also introduce a more sophisticated stopping criterion called LSC (Length Stopping Criterion) based on Theorem~\ref{T:bound} such that if $\sum_{l=1}^{s-1}w(\sigma(l),\sigma(l+1))+\overline{w}(\sigma(s),\sigma(1)) > \max_{t \in T}\{d(t)\}$, then the search is stopped and backtracked. We introduce also an \emph{a priori} check (IFC, initial Forward Checking): before starting the search, we consider each vertex as the root node of the search tree and we apply the forward checking. If at least one domain is empty, the algorithm returns failure. Otherwise, the tree search is started.

Finally, we introduce some heuristic criteria for choosing the next vertex to expand in Step~8 of Algorithm~\ref{A:recursivecall}: lexicographic ($h_l$), random with uniform probability distribution ($h_r$), maximum and minimum number of incident arcs ($h_{max~a}$ and $h_{min~a}$), less visited ($h_{min~v}$), and maximum and minimum penetration time ($h_{max~d}$ and $h_{min~d}$). For all the ordering criteria except $h_r$, we introduce a criterion for breaking ties (RTB, Random Tie-Break) that selects a vertex with uniform probability. We used the previous heuristics also for selecting the initial node of the search tree in Step~1 of Algorithm~\ref{A:init}.

\subsection{Experimental Results}
\label{S:deterministic:experiments}
In this section, we experimentally evaluate the performance of our algorithm in producing a deterministic equilibrium strategy or in returning a failure. We developed a random generator of graphs $G'$ with parameters $n$ (number of vertices, corresponding to targets) and $m$ (number of arcs) and working as follows. Given two values $n$ and $m$, firstly a random connected graph with $n$ vertices is produced, then $m-n$ arcs are added. All the arcs' weights are set equal to $1$ (it can be easily shown that this is the worst case for computational complexity). Values $d({k})$ are uniformly drawn from the interval $[ \min_{i,j}{ \{ \overline{w}(i,j) + \overline{w}(j,i)\} }, 2 n^{2} \max_{i,j}{\overline{w}(i,j)} ]$, where $\overline{w}(i,j)$ is the length of the shortest path between vertices ${i}$ and ${j}$. The lower bound of the interval comes from the consideration that settings with $d({k})< \min_{i,j}{ \{ \overline{w}(i,j) + \overline{w}(j,i) \} }$ are infeasible and our algorithm immediately detects that unfeasibility (by IFC). The upper bound is justified by considering that if a problem is feasible then it always admits a solution shorter than $2 n^{2} \max_{i,j}\{\overline{w}(i,j)\}$. Graphs differ from each other in the topology and in the penetration times of the vertices. This program and those implementing our algorithms have been coded in C and executed on a Linux (2.6.24 kernel) computer equipped with a DUAL QUAD CORE Intel XEON 2.33 GHz CPU, 8 GB RAM, and 4 MB cache.

For each ordering criterion (i.e., $h_l$, $h_r$, $h_{max~a}$, $h_{min~a}$, $h_{min~v}$, $h_{max~d}$, $h_{min~d}$) with and without LSC and IFC and for each $n\in\{3,4,5,6,7,8,100,250,500\}$ we produce 1000 patrolling settings with $m$ uniformly drawn from the interval $[n, (n-1)n]$ (if $m<n$ the graph is not connected, if $m>(n-1)n$ at least a pair of vertices is connected by more than one arc). We evaluate the percentage of terminations of the algorithm within 10 minutes and, in the case of termination (either with a solution or with a failure), the computational time. Since, as discussed in the Section~\ref{S:deterministic:problem}, our approach is aimed at finding a solution and not the optimal solution according to a given metric (e.g., the cycle length), we do not measure the quality of a solution.

\begin{table}[htbp]
   \centering
   \begin{scriptsize}
   \begin{tabular}{|c|c|c|c|c|c|c|c|c|c|c|c|}
      \hline
   & $n$     & $3$   & $4$     & $5$  & $6$ & $7$ & $8$ & $100$ & $250$ & $500$\\  \hline\hline
$h_{min~v}$    & \%             & $100$         & $100$         & $100$         & $99.8$     & $99.6$     & $99.5$      &   $98.9$    &   $96.6$     &   $90.2$  \\  \cline{2-11}
$RTB$    & time [s]    & $<0.01$     & $<0.01$      & $<0.01$        & $0.32$        & $0.10$       & $0.05$        &   $0.16$      &   $0.87$       &   $5.50$\\  \cline{2-11}
$LSC$    & ~dev [s]       & $<0.01$     &  $<0.01$     & $<0.01$        & $5.17$     & $1.78$     & $0.96$       &   $3.52$      &   $14.47$     &   $30.28$\\  \cline{2-11}
$IFC$ & max [s]             & $<0.01$     &  $<0.01$      & $<0.01$        & $98.00$         & $35.00$        & $19.00$             &   $78.26$    &   $316.9$     &   $413.94$\\  \cline{2-11}
& ~min [s]             & $<0.01$     &  $<0.01$      & $<0.01$     & $<0.01$     & $<0.01$     & $<0.01$      &   $<0.01$     &  $0.01$       &   $0.07$\\  \hline   \hline
$h_{r}$    & \%             & $100$         & $100$         & $100$         & $98.5$     & $97.5$     & $96.5$ & $95.1$ & $55.1$ & $9.8$ \\  \cline{2-11}
$LSC$    & time [s]    & $<0.01$     & $<0.01$      & $0.11$        & $0.09$        & $0.16$       & $0.02$ & $1.34$ & $2.52$ & $4.66$ \\  \cline{2-11}
$IFC$    & ~dev [s]       & $<0.01$     &  $<0.01$     & $1.64$        & $1.70$     & $1.73$     & $0.18$ & $6.19$ & $16.75$ & $51.62$ \\  \cline{2-11}
& max [s]             & $<0.01$     &  $<0.01$      & $32.00$        & $33.00$         & $24.00$        & $2.00$ & $93.36$ & $513.66$ & $590.87$ \\  \cline{2-11}
& ~min [s]             & $<0.01$     &  $<0.01$      & $<0.01$     & $<0.01$     & $<0.01$     & $<0.01$ & $<0.01$ & $0.01$& $0.07$ \\  \hline   \hline
$h_{r}$    & \%             & $100$         & $100$         & $99.0$         & $97.2$     & $96.7$     & $95.5$ & $94.0$ & $53.0$ & $8.9$ \\  \cline{2-11}
$IFC$    & time [s]    & $<0.01$     & $0.44$      & $3.65$        & $0.14$        & $0.26$       & $0.01$ & $7.12$ & $3.41$ & $5.94$ \\  \cline{2-11}
    & ~dev [s]       & $<0.01$     &  $8.68$     & $38.89$        & $2.24$     & $2.36$     & $0.16$ & $39.32$ & $18.02$ & $55.14$ \\  \cline{2-11}
& max [s]             & $<0.01$     &  $173.55$      & $594.10$        & $43.03$         & $31.86$        & $2.09$ & $561.95$& $501.72$ & $582.77$ \\  \cline{2-11}
& ~min [s]             & $<0.01$     &  $<0.01$      & $<0.01$     & $<0.01$     & $<0.01$     & $<0.01$ & $<0.01$& $0.01$& $0.07$ \\  \hline   \hline
$h_{min~v}$    & \%             & $100$         & $100$         & $100$         & $96.7$     & $96.0$     & $95.5$ & $95.0$ & $93.3$ & $86.2$\\  \cline{2-11}
$RTB$    & time [s]    & $<0.01$     & $<0.01$      & $0.34$        & $2.98$        & $0.16$       & $0.01$ & $0.30$& $1.00$ & $6.19$ \\  \cline{2-11}
$LSC$    & ~dev [s]       & $<0.01$     &  $<0.01$     & $6.29$        & $33.77$     & $2.24$     & $0.11$ & $6.50$& $15.32$ & $35.77$\\  \cline{2-11}
& max [s]             & $<0.01$     &  $<0.01$      & $125.03$        & $519.75$         & $42.22$        & $2.41$ & $145.22$ & $366.42$ & $498.04$\\  \cline{2-11}
& ~min [s]             & $<0.01$     &  $<0.01$      & $<0.01$     & $<0.01$     & $<0.01$     & $<0.01$ & $<0.01$ & $0.01$& $0.07$ \\  \hline   \hline
$h_{r}$    & \%             & $100$         & $100$         & $100$         & $95.4$     & $93.9$     & $92.5$ & $91.2$ &$52.4$ &$7.7$ \\  \cline{2-11}
$LSC$    & time [s]    & $<0.01$     & $<0.01$      & $0.79$        & $3.04$        & $0.30$       & $0.03$ & $7.16$& $3.48$ & $5.83$ \\  \cline{2-11}
    & ~dev [s]       & $<0.01$     &  $<0.01$     & $13.58$        & $24.32$     & $2.77$     & $0.21$ & $39.53$& $18.46$ & $55.65$ \\  \cline{2-11}
& max [s]             & $<0.01$     &  $<0.01$      & $270.03$        & $303.72$         & $41.53$        & $2.83$ & $566.04$& $531.64$ & $596.42$ \\  \cline{2-11}
& ~min [s]             & $<0.01$     &  $<0.01$      & $<0.01$     & $<0.01$     & $<0.01$     & $<0.01$ & $0.02$& $0.01$& $0.07$ \\  \hline   \hline
$h_{r}$    & \%             & $100$         & $100$         & $98.7$         & $94.2$     & $93.0$     & $91.8$ & $90.3$ & $51.0$ & $7.1$ \\  \cline{2-11}
    & time [s]    & $<0.01$     & $7.45$      & $2.45$        & $4.78$        & $1.38$       & $0.14$ & $1.37$& $3.74$ & $6.18$ \\  \cline{2-11}
    & ~dev [s]       & $<0.01$     &  $55.45$     & $28.61$        & $42.13$     & $9.96$     & $1.03$ & $6.28$& $18.45$ & $56.80$ \\  \cline{2-11}
& max [s]             & $<0.01$     &  $556.92$      & $506.72$        & $496.84$         & $140.31$        & $12.86$ & $93.26$& $516.72$ & $576.52$ \\  \cline{2-11}
& ~min [s]             & $<0.01$     &  $<0.01$      & $<0.01$     & $<0.01$     & $<0.01$     & $<0.01$ & $0.02$& $0.01$& $0.07$ \\  \hline   \hline
$h_{l}$    & \%             & $100$         & $99.2$         & $91.0$         & $81.1$     & $75.3$     & $69.0$ & $3.9$ & $2.3$ & $1.5$\\  \cline{2-11}
$LSC$    & time [s]    & $<0.01$     & $7.45$      & $2.45$        & $4.78$        & $1.38$       & $0.14$ & $0.10$ & $0.01$& $0.07$\\  \cline{2-11}
$IFC$    & ~dev [s]       & $<0.01$     &  $55.45$     & $28.61$        & $42.13$     & $9.96$     & $1.03$ & $<0.01$ & $<0.01$& $<0.01$\\  \cline{2-11}
& max [s]             & $<0.01$     &  $548.41$      & $505.74$        & $497.46$         & $140.11$        & $12.01$ & $<0.01$ & $0.01$ & $0.07$ \\  \cline{2-11}
& ~min [s]             & $<0.01$     &  $<0.01$      & $<0.01$     & $<0.01$     & $<0.01$     & $<0.01$ & $<0.01$& $0.01$& $0.07$\\  \hline   \hline
$h_{l}$    & \%             & $100$         & $99.2$         & $88.0$         & $78.0$     & $71.7$     & $65.0$ & $0.0$ & $0.0$& $0.0$\\  \cline{2-11}
    & time [s]    & $<0.01$     & $7.42$      & $2.61$        & $5.12$        & $1.61$       & $0.20$ & $-$ & $-$& $-$\\  \cline{2-11}
    & ~dev [s]       & $<0.01$     &  $55.23$     & $28.66$        & $42.65$     & $10.57$     & $1.29$ & $-$ & $-$ & $-$ \\  \cline{2-11}
& max [s]             & $<0.01$     &  $548.41$      & $505.74$        & $497.46$         & $140.11$        & $12.01$ & $-$ & $-$ & $-$ \\  \cline{2-11}
& ~min [s]             & $<0.01$     &  $<0.01$      & $<0.01$     & $<0.01$     & $<0.01$     & $<0.01$ & $-$ & $-$ & $-$ \\  \hline
\end{tabular}
  \end{scriptsize}
   \caption{Experimental results for different algorithm configurations.}
   \label{table:computationaltimes2}
\end{table}

The most significant experimental results we obtained are reported in Table~\ref{table:computationaltimes2}: the numbers reported in the table are averaged over the 1000 settings that have been generated as discussed above. The table shows the termination percentage and, for terminated runs, the average time, its standard deviation, and the maximum and minimum observed times. The first remark is that, for all the algorithm configurations, the averaged computational time is reasonably short also for large settings. Instead, the termination percentage differs very much in different configurations.  In this sense, the behavior of the proposed algorithm resembles that of many constraint programming algorithms, whose termination time is usually either very short (when a solution is found) or the algorithms do not terminate within the deadline. The random generation of graphs explains the data relative to the maximum computational time: some cases are harder than the average and require a lot of time to be solved (in practice, they both reduce the percentage of termination and increase the computational time). These hard cases, which represent outliers of the population of graphs, are characterized by complicated topologies or oddly-distributed relative deadlines.

Here are some comments regarding other issues.

\begin{description}
\item[Ordering criteria] The best ordering criterion is $h_{min~v}$ with RTB. We omit the experimental results with $h_{max~a}$, $h_{min~a}$, $h_{max~d}$, and $h_{min~d}$, since they are very similar to those obtained with $h_l$. The criterion $h_{min~v}$ with RTB leads the algorithm to terminate with a percentage close to $h_r$ for small values of $n$ and about 80\% larger for large values of $n$. Instead, $h_l$ provides very bad performance, especially for large values of $n$, when the algorithm terminates with percentages close to 0\%.
\item[LSC] The improved stopping criterion allows the algorithm to increase the termination percentage by a value between 0\% and 2\%, without distinguishable effects on the computational time. This improvement  depends on the configuration of the algorithm since it affects the construction of the search tree.
\item[IFC] This criterion allows the algorithm to increase the termination percentage by a value between 1\% and 4\%, reducing the computational time (since all the non-feasible linear settings are solved with a negligible computational time). This improvement does not depend on the configuration of the algorithm since it does not affect the search, working before it.
\end{description}
Therefore the best algorithm configuration is $h_{min~v}$ with RTB, LSC, and IFC. With this configuration, the results are good: the termination percentage is very high also for large settings, like those with $500$ vertices, and the corresponding average computational time, about $5.5$~s, is reasonably short. We notice that in some practical indoor settings the number of vertices is in the range $\{24,\ldots,100\}$, see, e.g.,~\cite{kolling2008}.

\section{Finding Non-Deterministic Equilibrium Patrolling Strategies}
\label{section:nondeterministic}

In this section, we consider the problem of finding non-deterministic equilibrium patrolling strategies for the game formulated in Section~\ref{subsez:gamemodel}. We recall that, according to the proposed approach, given a patrolling setting, we first look for a deterministic equilibrium strategy that, if found, makes attacking a target not rational for the intruder. If a deterministic equilibrium strategy cannot be found, we look for a non-deterministic equilibrium strategy, following the algorithm presented in this section to find leader-follower equilibria. In what follows, we survey the main related works on the computation of leader-follower equilibria (Section~\ref{subsection:nondeterministicstateoftheart}), we present our solving algorithm (Section~\ref{S:nondeterministic:solving}), and we experimentally evaluate it (Section~\ref{section:nondeterministic:experimental}).

\subsection{Related Works}
\label{subsection:nondeterministicstateoftheart}

The literature presents some works on the computation of leader-follower equilibria that are based on operational research techniques. Basically, the computation of a leader-follower equilibrium can be formulated as a multi-level optimization problem~\cite{Colson05bilevel}, where each level corresponds to a specific player. In our case, the first optimization level is associated to the patroller and the second one to the intruder. When both the optimization problems (at the first  and second level) are linear, the algorithmic game theory literature provides assessed techniques for their solution~\cite{sandholm}. When, instead, at least an optimization problem (at the first or second level) is not linear, as it turns out to be in our case, works in literature provide more a collection of examples than a coherent set of results. We briefly review these works.

The main result on which other algorithms for finding leader-follower equilibria are based is by von~Stengel and Zamir~\cite{commitmentbased}: at the equilibrium, the follower always employs pure strategies, playing the best response for the strategy the leader committed to. (We recall that, at the equilibrium, the follower, when more equivalent best responses are available,  must choose the one that maximizes the leader's expected utility.) 

Two are the main works addressing the situation in which both the levels of the optimization problem are linear. The first one is by Conitzer and Sandholm~\cite{sandholm}. The authors formulate the problem of finding a leader-follower equilibrium as a multi-linear programming problem (Multi-LP). The basic idea is, for each action $a$ of the follower, to compute the (mixed) strategy of the leader $\sigma_l$ that maximizes the leader's expected utility under the constraint that $a$ is a follower's best response. Among all such strategies $\sigma_l$s, the leader will choose the one that maximizes its expected utility. With this method, there are as many optimization problems as pure strategies of the follower and each single optimization problem is linear. Conitzer and Sandholm show in~\cite{sandholm} that the problem of computing mixed-strategy leader-follower equilibria in two-player complete-information games is polynomial in the size of the game. The same result holds also when the leader's type is uncertain. Instead, when the follower's type is uncertain the problem is NP-hard. The second work is by Paruchuri \emph{et al.}~\cite{paruchuri2008}, in which the authors propose an alternative mathematical programming formulation based on mixed integer linear programming (MILP) that is shown to be more efficient with multiple follower's types with respect to the Multi-LP formulation of~\cite{sandholm}. The basic idea behind this alternative approach is to represent the follower's actions as binary variables where value 1 means that the corresponding action is played and value 0 means that it is not. 

As said, when the optimization problems involved in the determination of leader-follower equilibria are non-linear, no assessed technique is available. Non-linear optimization problems have been studied mainly when the non-linearity lays in the second level and the optimization problem is convex and regular. In these cases, the Karush-Khun-Tucker~\cite{Bazaraa2006nonlinear} theorem is employed to linearize the optimization problem. In our case, we cannot apply this method because, as discussed later, our problem is not convex. Alternative methods are based on non-exact linearization and produce approximate solutions~\cite{Bazaraa2006nonlinear}. 

Computing equilibria in games in practical settings usually requires a large computational effort. A simple approach to save computational time is the reduction of the search space. Customarily, in game theory, this is accomplished by exploiting the concept of \emph{dominance}: action $a$ of player $i$ is said to be (weakly) dominated by another action $a'$ of player $i$ if player $i$'s expected utility from playing action $a$ is never larger than expected utility from playing action $a'$ (independently of the actions of the other players). Dominated actions can be safely removed, since they will be never played by rational agents. Usually, removing dominated actions is computationally inexpensive and drastically reduces the search space, improving the computational efficiency of the solving algorithm~\cite{shoham-book}.

\subsection{Solving Algorithm}
\label{S:nondeterministic:solving}

In this section, we overview the proposed solving algorithm (Section~\ref{S:nondeterministic:solving:outline}), we describe in detail its steps (Section~\ref{S:nondeterministic:solving:domination} and Section~\ref{subsez:programming}), we provide a specific formulation for strictly competitive settings (Section~\ref{section:nondeterministic:algorithm:strictlycompetitive}), and we discuss some theoretical properties of the algorithm (Section~\ref{section:nondeterministic:algorithm:properties}).

\subsubsection{Algorithm Overview}
\label{S:nondeterministic:solving:outline}

Our algorithm works on the original graph $G$ as defined in Section~2.1. We shall show in Section~\ref{section:nondeterministic:algorithm:properties} that a reduction to a graph $G'$ composed only of target vertices (as done in Section~3) cannot be applied when searching for a non-deterministic equilibrium strategy. The algorithm depends on the value of $l$, namely on the finite length of actions' history. In this paper, we provide the algorithm for the case $l=1$. This is because, formulations with $l=0$ are applicable only to environments with fully connected topologies (as discussed in Section~\ref{subsez:stackelbger} and further detailed in Section~\ref{section:nondeterministic:algorithm:properties}) and formulations with $l>1$, as we shall discuss below, can be obtained by extending the case with $l=1$.  Our algorithm develops in three steps as follows.
\begin{enumerate}
\item The first step removes the dominated actions (of the patroller and of the intruder) that the agents will never play. 
\item The aim of this step is to check whether or not there exists a patroller's strategy such that intruder's action $\textit{stay-out}$ is a best response. If there exists such a strategy, it is the optimal patrolling strategy. Otherwise, the algorithm passes to the third step.
\item The aim of this step is to compute the equilibrium strategy for the patroller under the assumption that the intruder will not make $\textit{stay-out}$.
\end{enumerate}

\subsubsection{Removing Dominated Actions}
\label{S:nondeterministic:solving:domination}

The first step of our algorithm is the removal of agents' dominated actions. We remark that, to the best of our knowledge, ours is the first attempt to apply dominated action removal to patrolling problems. This step splits into two phases: at first we remove the patroller's dominated actions and, subsequently, we remove the intruder's dominated actions.

\begin{figure}[htb]
\centering
\includegraphics[width=0.6\textwidth]{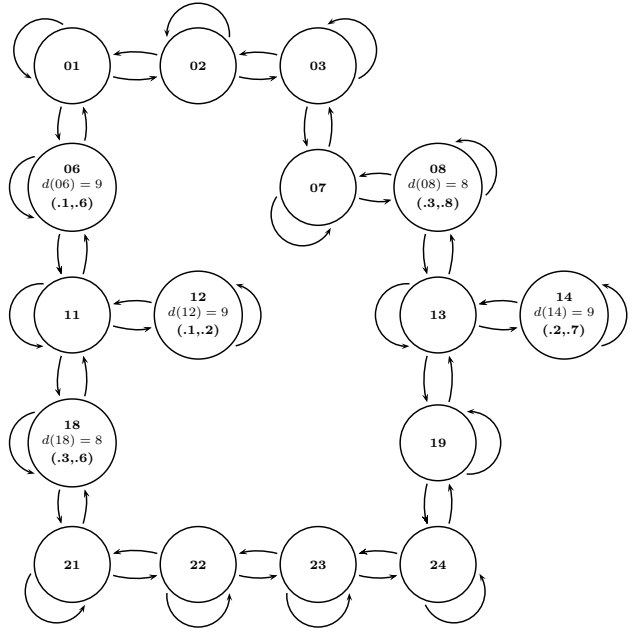}
\caption{Graph $G_{r}$ for the patrolling setting of Figure~\ref{F:graforunningexample}, obtained by removing the vertices corresponding to the patroller's dominated actions.}
\label{fig:nondominatedpatroller}
\end{figure}

We focus on the patroller's side. We recall that the actions of the patroller are of the form $\textit{move}(i)$. This means that, if action $\textit{move}(i)$ is dominated, the patroller will never visit vertex $i$ and then such vertex can be eliminated from the patrolling problem. Therefore, by discarding patroller's dominated actions, we obtain a reduction of the graph $G$. Basically, we can remove any vertex $i$ (and all its ingoing and outgoing arcs) such that the corresponding action $\textit{move}(i)$ is dominated by another action $\textit{move}(j)$ (with $j \neq i$),  independently of the intruder's strategy, in the following way.
\begin{enumerate}
\item For each target $t$, we remove from $G$ any vertex $i$ such that the shortest distance between $i$ and $t$ is strictly larger than the penetration time $d(t)$. A rational patroller will never visit such vertices. Indeed, if it visits them, then the intruder will attack target $t$ being sure not to be captured in the next $d(t)$ turns. If, after this removal, the graph is not connected, then there is not any non-deterministic equilibrium strategy that can cover all the targets. In this case, as discussed in Section~2.4, multiple robots should be adopted.
\item After having reduced the graph $G$ as described above, we compute the shortest paths for any pair of targets $t, t'$ and we remove all the vertices that do not belong to any shortest path. The idea is that if a vertex $z$ is not on any shortest path and a strategy $\sigma_{\mathbf{p}}$ prescribes that the patroller can make action $\textit{move}(z)$ with strictly positive probability, then it can be easily observed that, if the patroller does not make such action, it cannot decrease its expected utility. Indeed, the probability to reach any target $t$ within $d(t)$ turns cannot decrease since visiting $z$ would introduce an unnecessary temporal cost. In the case there are multiple shortest paths between two targets, we keep all of them (in some specific cases, we could select a particular shortest path as we showed in~\cite{BasilicoGattiAmigoniAAMAS2009}).
\end{enumerate}
We call $G_r=(V_r,A_r,T,v,d)$ the reduced graph produced as prescribed by the two above steps. Graph $G_r$ can be obtained in linear time in the size $n$ (number of vertices) of the patrolling setting, given the shortest paths. We recall that these shortest paths have been already computed (by applying Dijkstra's algorithm) when our algorithm searched for the deterministic equilibrium strategy. We report in Figure~\ref{fig:nondominatedpatroller} the graph $G_{r}$ for our running example of Figure~\ref{F:graforunningexample} after having removed the vertices corresponding to the dominated actions $\textit{move}(i)$ of the patroller. The dominated actions are: $\textit{move}(i)$ with $i\in\{04,05,09,10,15,16,17,20,25,26,27,28,29\}$.

We now focus on the intruder's actions. We recall that the actions of the intruder are of the form $\textit{enter-when}(t,h)$, where $t$ is a target and $h$ is a history, and $\textit{stay-out}$. Considering $l=1$, actions $\textit{enter-when}(t,h)$ reduce to $\textit{enter-when}(t,i)$, with the meaning that the intruder enters target $t$ after it observed the patroller in vertex $i$. We consider only dominance between actions $\textit{enter-when}(t,i)$. This means that, if action $\textit{enter-when}(t,i)$ is dominated, the intruder will never attack $t$ after having observed the patroller in $i$ and then such action can be discarded. We remove intruder's dominated actions in the following way.

Fixed a target $t$, $\textit{enter-when}(t,i)$ is dominated by another intruder's action $\textit{enter-when}(t,i')$ if the patroller, when covering every path starting from $i'$ and arriving at $t$ before $d(t)$ turns, must always visit vertex $i$. The basic idea is: the intruder's expected utility for action $\textit{enter-when}(t,i)$ depends on the probability that such action will lead to a successful intrusion. This probability is related to the probability that the patroller reaches $t$ within $d(t)$ turns from its current vertex $i$. It can be easily observed that, in the case the patroller, starting from $i'$, must always visit $i$ to reach $t$ within $d(t)$ turns, the probability that the patroller reaches $t$ starting from $i$ within $d(t)$ turns is not smaller than the probability  that the patroller reaches $t$ from $i'$ within $d(t)$ turns (it is a trivial application of Markov chains). Therefore, the  expected utility (for the intruder) of $\textit{enter-when}(t,i)$ is not larger than the expected utility of $\textit{enter-when}(t,i')$. This holds independently of the patroller's strategy. It can be easily observed, instead, that, given two different targets $t_1$ and $t_2$, actions $\textit{enter-when}(t_1,i)$ and $\textit{enter-when}(t_2,i')$ may not have any dominance relationship for any possible $i$ and $i'$ independently of the patroller's strategy. 

Consider the example reported in Figure~\ref{fig:nondominatedpatroller}. Consider target $08$ and vertices $01$ and $02$. The probability that the patroller reaches target $08$ within $d(08)=8$ turns starting from $01$ is not larger than the probability that the patroller reaches target $08$ within $d(08)=8$ turns starting from $02$. Indeed, when the patroller starts from vertex $01$, it must always pass through vertex $02$ to reach $08$ by 8 turns. Therefore, the intruder will prefer to play $\textit{enter-when}(08,01)$ rather than $\textit{enter-when}(08,02)$ independently of the patroller's strategy. This shows why, in this example, $\textit{enter-when}(08,01)$ dominates $\textit{enter-when}(08,02)$ and this last action can be removed.

\begin{figure}[htb]
\centering
\includegraphics[width=0.5\textwidth]{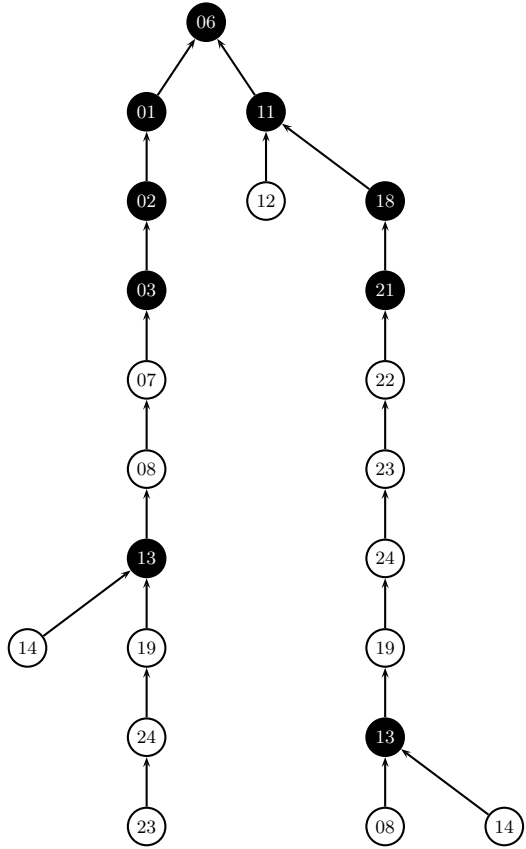}
\caption{Search tree for finding dominated actions for target 06 of Figure~\ref{fig:nondominatedpatroller}.}
\label{fig:dominance}
\end{figure}

We call $V_{t} \subseteq V_r$ the subset of vertices that satisfy the following condition: for every $i\in V_t$ action $\textit{enter-when}(t,i)$ is not dominated. Therefore, $V_t$ provides a compact representation of the non-dominated actions. Set $V_t$ can be found by resorting  to tree search techniques in the following way. We denote by $q$ a node in the search tree and by $\eta(q)$ the vertex corresponding to $q$. We build a tree of paths where, called $q_0$ the root, $\eta(q_0)=t$ and a node $q''$ is a successor of $q'$ if and only if  $q'' \in \{q \textnormal{ s.t. } a(\eta(q'),\eta(q)) = 1\}$ and $q'' \neq q'$ and $q'' \neq father(q')$. The maximum depth of the tree is $d(t)$. That is, we consider all the paths not longer than $d(t)$ and with $t$ as first vertex. Figure~\ref{fig:dominance} reports the tree of paths generated for $t=06$. Given a tree of paths so built, an action $\textit{enter-when}(t,i)$ is dominated when there exists a vertex $i'$ such that for each $q$ with $\eta(q)=i$ there is a $q'$ with $\eta(q')=i'$ and $father(q')=q$. In this case, $i \not\in V_{t}$. In Figure~\ref{fig:dominance}, black nodes denote vertices $i$ such that actions $\textit{enter-when}(06,i)$ are dominated; for example, action $\textit{enter-when}(06,13)$ is dominated since every occurrence of vertex 13 in the search tree has a node with vertex 14 as child. The computational complexity of removing intruder's dominated actions is $O(|T| b^{\max_{t \in T}\{ d(t)\}})$ where $T$ is the set of targets, $b$ is the largest number of outbound  arcs from a target, and $\max_{t \in T}\{ d(t)\}$ is the largest penetration time.

To summarize, the results produced by the first step of our algorithm are:
\begin{itemize}
\item a reduced patrolling setting $G_r$, obtained from $G$ by removing some vertices and their corresponding arcs that represent patroller's dominated actions,
\item for each $t \in T$ a subset of vertices $V_t$, containing the vertices such that every action $\textit{enter-when}(t,i)$ with $i \in V_t$ is not dominated.
\end{itemize}

\subsubsection{Mathematical Programming Formulation}
\label{subsez:programming}

In this section, we illustrate the second and the third step of our algorithm. We present them together since they are conceptually similar, being both based on mathematical programming. The solution of the mathematical programming problem is the non-deterministic equilibrium patrolling strategy we are looking for. More precisely, we provide two mathematical programming formulations, one for each step of the algorithm, whose solution can be obtained using optimization software tools, e.g.,~\cite{snopt}. As before, we provide our mathematical formulations when $|h|=l=1$. With $l=1$, the Markov hypothesis holds and the patroller's strategy $\{ \alpha_{\langle i \rangle, \textit{move}(j)} \}$ introduced in Section~\ref{subsez:stackelbger} can be compactly represented by $\{ \alpha_{i,j} \}\;\forall i,j \in V_r$, where each $\alpha_{i,j}$ denotes the probability for the patroller to move from vertex $i$ to vertex $j$ (namely, to take action $\textit{move}(j)$ while in $i$). Our formulation is inspired by~\cite{sandholm} and introduces some non-linearities~\cite{Bazaraa2006nonlinear}. More precisely, given a pure strategy of the intruder $\sigma_{\mathbf{i}}=a$, the maximization of the patroller's expected utility is linear in the objective and bilinear in the constraints. The non-linearity is due to the symmetries of our game model that are introduced by the Markov hypothesis: we need to constrain behavioral strategies\footnote{We recall that a behavioral strategy of an agent in a given decision node is the strategy conditioned by the agent being at such node.} to be equal in the same state, i.e., $\alpha_{i,j}$ is fixed for all the decision nodes for which the patroller's current vertex is $i$. This non-linearity forces us to avoid a mixed integer formulation (as in~\cite{paruchuri2008}) that, in our case, would be a mixed integer non-linear problem whose efficient solution is still an open issue in the operational research field. In order to have the minimal non-linear degree (i.e., quadratic degree) we took inspiration form the sequence-form proposed in~\cite{koller}.

We now present the details of the second step of our algorithm where  we check whether there exists at least one patroller's strategy $\sigma_{\mathbf{p}}$ such that $\textit{stay-out}$ is a best response for the intruder. If such a strategy exists, then the patroller will follow it, being its utility maximum when the intruder abstains from the intrusion (recall the utility definition of Section~\ref{subsez:gamemodel}). This step is formulated as a bilinear feasibility problem in which $\alpha_{i,j}$s are the decision variables. We denote by  $V_{r}\setminus i$ the set obtained by removing element $i$ from set $V_{r}$ and by $\gamma_{i,j}^{w,t}$ the probability that the patroller reaches vertex $j$ in $w$ steps, starting from vertex $i$ and not sensing (i.e., not passing through) target $t$. The feasibility problem is the following:

\begin{scriptsize}
\begin{eqnarray}
\alpha_{i,j} \ge 0 &\forall i,j \in V_r \label{con:feasibility1}\\
\sum_{j \in V_r}\alpha_{i,j} = 1&\forall i \in V_r \label{con:feasibility2}\\
\alpha_{i,j} \le a_{r}(i,j)&\forall i,j \in V_r \label{con:feasibility3}\\
\gamma_{i,j}^{1,t} =\alpha_{i,j} & \forall t\in T, i,j \in V_r, j \neq t \label{con:feasibility4}\\
\gamma^{w,t}_{i,j} = \sum_{x \in V_r\setminus t}\left(\gamma^{w-1,t}_{i,x}\alpha_{x,j}\right)& \forall w \in \{2,\ldots,d(t)\},\forall t\in T,i,j \in V_r, j \neq t \label{con:feasibility5}\\
\begin{split}u_{\mathbf{i}}(\textit{intruder-capture}) \left(1-\sum_{i \in V_r \setminus t}\gamma^{d(t),t}_{z,i}\right) +\\+ u_{\mathbf{i}}(\textit{penetration-$t$})\sum_{i \in V_r \setminus t}\gamma^{d(t),t}_{z,i}\leq0\end{split} &\forall t \in T, z \in V_t \label{con:feasibility6}
\end{eqnarray}
\end{scriptsize}

\noindent Constraints~(\ref{con:feasibility1})-(\ref{con:feasibility2}) express that probabilities $\alpha_{i,j}$s are well defined; constraints~(\ref{con:feasibility3}) express that the patroller can only move between two adjacent vertices; constraints~(\ref{con:feasibility4})-(\ref{con:feasibility5}) express the Markov hypothesis over the patroller's decision policy; constraints~(\ref{con:feasibility6}) express that no action $\textit{enter-when}(t,z)$ gives to the intruder an expected utility larger than that of $\textit{stay-out}$. The non-linearity is due to constraints~(\ref{con:feasibility5}). If the above problem admits a solution, the resulting $\alpha_{i,j}$s are the optimal patrolling strategy. We notice that, due to constraints~(\ref{con:feasibility5}), the above feasibility problem is not convex and we cannot linearize it by applying Karush-Khun-Tucker theorem (see Section~\ref{subsection:nondeterministicstateoftheart}). When no dominated action can be removed, the problem presents $O(mn^2\max_{t \in T}\{d(t)\})$ variables and constraints (where $n$ is the number of vertices in $G$ and $m$ is the number of targets). In practical settings, removing dominated actions drastically reduces the number of variables and constraints, as we shall show in Section~\ref{section:nondeterministic:experimental}. 

We now discuss the relationships between this second step of our algorithm to compute non-deterministic equilibrium strategies and the algorithm presented in Section~3 to compute deterministic equilibrium strategies. The similarities between these two algorithms are due to the fact that they both produce patrolling strategies (if they exist) such that the intruder's best response is $\textit{stay-out}$. Anyway, their scope is different and they are both necessary. Indeed, if there exists a deterministic equilibrium strategy with $|h|=l>1$, the above mathematical programming problem fails in finding it, since it is not Markovian, and such strategy can be found only  applying the algorithm of Section~3. On the other side, when no deterministic equilibrium strategy exists, there could be a non-deterministic equilibrium strategy such that the intruder's best response is $\textit{stay-out}$. Obviously, an extension of the above formulation with $l$ very large would make the algorithm for the computation of deterministic equilibrium strategies unnecessary. However, such a formulation is expected to be hardly solvable in practice.

When the above feasibility problem does not admit any solution (i.e., there is not any patroller's strategy such that $\textit{stay-out}$ is a best response for the intruder), we pass to the third step of the algorithm. In this step, we find the best response of the intruder such that the patroller can maximize its expected utility. This problem is formulated as a multi-bilinear programming problem. The single bilinear problem, in which $\textit{enter-when}(s,q)$ with $s\in T$ and $q\in V_s$ is assumed to be the intruder's best response, is defined as:

\begin{scriptsize}
\begin{equation*}
\max\qquad u_\mathbf{p}(\textit{penetration-}s) \sum_{i \in V_r \setminus s}\gamma^{d(s),s}_{q,i} + u_\mathbf{p}(\textit{intruder-capture})\left(1 - \sum_{i \in V_r \setminus s}\gamma^{d(s),s}_{q,i}\right)
\end{equation*}
s.t.
\begin{eqnarray*}
\textnormal{constraints (\ref{con:feasibility1})-(\ref{con:feasibility5})}
\end{eqnarray*}
\begin{eqnarray}
\begin{split} u_{\mathbf{i}}(\textit{intruder-capture})  \left(\sum_{i \in V_r \setminus t}\gamma^{d(t),t}_{z,i}-\sum_{i \in V_r \setminus s}\gamma^{d(s),s}_{q,i}\right) +\\+ u_\mathbf{i}(\textit{penetration-}s)\sum_{i \in V_r \setminus s}\gamma^{d(s),s}_{q,i}  - u_\mathbf{i}(\textit{penetration-}t)\sum_{i \in V_r \setminus t}\gamma^{d(t),t}_{z,i}\geq 0\end{split} &\forall t\in T, z \in V_t \label{cons:multibilinear}
\end{eqnarray}
\end{scriptsize}

\noindent The objective function maximizes the patroller's expected utility. Constraints~(\ref{cons:multibilinear}) express that no action $\textit{enter-when}(t,z)$ gives a larger value to the intruder than action $\textit{enter-when}(s,q)$. When no dominated action can be discarded, we can formulate $m\cdot n$ above problems for all the possible actions $\textit{enter-when}(s,q)$ with $s\in T, q\in V_r$. Practically, the number of bilinear problems to be considered is usually much smaller thanks to elimination of dominated actions. The size, in terms of variables and constraints, of each optimization problem is the same of the feasibility problem of the second step of our algorithm. If a bilinear problem is feasible, its solution is a set of probabilities $\{ \alpha_{i,j} \}$, that define a possible patrolling strategy. If the problem is unfeasible, then there is not any patroller's strategy such that $\textit{enter-when}(s,q)$ is a best response for the intruder. From all the solutions of feasible problems, we pick out the one that gives the patroller the maximum expected utility. As a final step, we need to check whether or not all the targets can be covered by the strategy. This can be easily accomplished by inspecting the randomized strategy. If all the targets can be covered with a strictly positive probability, the produced strategy is the optimal non-deterministic strategy for our patrolling setting that corresponds to the leader-follower equilibrium strategy. In the other case, there does not exist (with $l=1$) any non-deterministic equilibrium strategy that can cover all the targets. 

\begin{figure}[htb]
\begin{center}
\includegraphics[scale=0.75]{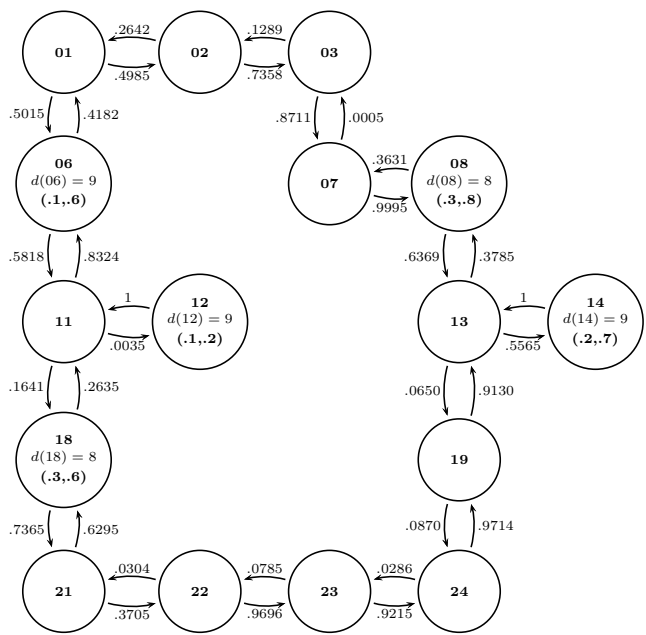}
\end{center}
\caption{Optimal patrolling strategy for Figure~\ref{fig:nondominatedpatroller}.}
\label{fig:strategia}
\end{figure}

We report in Figure~\ref{fig:strategia} the transition probabilities corresponding to the leader-follower equilibrium for the setting of Figure~\ref{fig:nondominatedpatroller}. The intruder's best response is $\textit{enter-when}(08,12)$.

It is worth a comment on how the size of the problem changes when $l=2$. In this case, variables $\alpha$s are defined as $\alpha_{h,z}$ with $h=\langle i,j\rangle$ representing the probability that the patroller moves to vertex $z$ after history  $h$ and variables $\gamma$ are defined as $\gamma_{h^0,h}^{w,t}$, with $h^0 = \langle i^0,j^0 \rangle$ and $h = \langle i,j \rangle$, representing the probability that the patroller reaches vertex $j$ from vertex $i$ in $w$ steps, starting after history $h^0$ and not passing through target $t$. The constraints with $l=2$ are similar to those with $l=1$. The problem presents $O(mn^4\max_{t \in T}\{d(t)\})$ variables and constraints.

\subsubsection{Improving Efficiency in Strictly Competitive Settings}
\label{section:nondeterministic:algorithm:strictlycompetitive}

Strictly competitive games are two-player games with specific constraints over players' preferences~\cite{fudenberg1991}. Call $i$ and $-i$ the two players. In strictly competitive games we have that for each possible pair of game outcomes $x,y$ if $u_i(x)>u_i(y)$, then $u_{-i}(x)<u_{-i}(y)$, and \emph{vice versa}. Therefore, in a strictly competitive game, if $u_i(x)=u_i(y)$ for two outcomes $x$ and $y$, then it necessarily must be that $u_{-i}(x)=u_{-i}(y)$. Zero sum games are a specific class of strictly competitive games. If a game is strictly competitive, its resolution is easier than when it is not. This is because, with strictly competitive games, von~Neuman's theorem assures that the maxmin,\footnote{The maxmin strategy of player $i$ is the strategy that maximizes $i$'s expected utility given that all the opponents happen to play the strategies which cause the greatest harm to $i$.} minmax,\footnote{The minmax strategy of player $i$ is the strategy that minimizes the $-i$'s expected utility.} and equilibrium (both Nash and leader-follower) strategies are the same~\cite{fudenberg1991}  and it turns out that computing maxmin and minmax strategies is much easier than computing an equilibrium strategy (Nash or refinements). As we shall show in this section, in a strictly competitive setting, the second and the third step of our algorithm collapse to a unique bilinear optimization problem of the size of the feasibility problem~(\ref{con:feasibility1})-(\ref{con:feasibility6}) of Section~\ref{subsez:programming}.

Our patrolling game can be studied as a strictly competitive game when for all targets $i,j\in T$ we have that, if $v_{\mathbf{p}}(i) \geq v_{\mathbf{p}}(j)$, then $v_{\mathbf{i}}(i) \geq v_{\mathbf{i}}(j)$, and \emph{vice versa}. Essentially we are requiring that both the patroller and the intruder have the same preference ordering over the targets. This assumption appears reasonable in a large number of practical settings, especially when values of the targets are common values. Rigorously speaking, when the above constraints hold, the game is not a strictly competitive game. This is because two outcomes (i.e., $intruder$-$capture$ and $no$-$attack$) provide the patroller with the same utility and the intruder with two (in general) different utilities (i.e., $-\epsilon$ and $0$, respectively). Anyway, we can temporarily discard the outcome $no$-$attack$, assuming that action $\textit{stay-out}$ will not be played by the intruder. We shall reconsider such action in the following. Without the outcome $no$-$attack$ and with the above constraints over the agents' valuations of the targets, the game is strictly competitive. We provide a mathematical programming formulation to find the patroller's minmax strategy, namely, the strategy that minimizes the intruder's expected utility and, the game being strictly competitive, maximizes the patroller's utility. We call $u$ the minmax value, i.e., the expected utility of the intruder. The mathematical programming formulation is the following:

\begin{scriptsize}
\[
\min u\\
\]
s.t.
\begin{eqnarray*}
\textnormal{constraints (\ref{con:feasibility1})-(\ref{con:feasibility5})}
\end{eqnarray*}
\begin{eqnarray}
u_{\mathbf{i}}(\textit{intruder-capture}) \left(1-\sum_{i \in V_r \setminus t}\gamma^{d(t),t}_{z,i}\right) + u_{\mathbf{i}}(\textit{penetration-$t$})\sum_{i \in V_r \setminus t}\gamma^{d(t),t}_{z,i}\leq u &\forall t \in T, z \in V_t \label{con:SCfeasibility6}\end{eqnarray}
\end{scriptsize}

Constraints~(\ref{con:SCfeasibility6}) assure that all the intruder's actions provide it with at most $u$ expected utility. It can be easily observed that the size, in terms of variables and constraints, of this mathematical programming problem is $O(mn^{2}\min_{t \in T}\{ d(t)\})$, which is the same of the feasibility problem of the second step of our algorithm for general, non-strictly competitive, games (see previous section).

Now, we reconsider action $\textit{stay-out}$ and its corresponding outcome $no$-$attack$. The basic idea is that the intruder will play $\textit{stay-out}$ if it pays better than any other action. Furthermore, the intruder knows that the utility of making $\textit{stay-out}$ independently of the patroller's strategy is 0. A simple approach is to solve the above mathematical programming problem and to compare $u$ with respect to 0: if $u<0$, then the intruder will play $\textit{stay-out}$, otherwise it will not. Therefore, in a strictly competitive setting, the second and the third step of our algorithm collapse to the resolution of a unique optimization problem plus a comparison between $u$ and $0$.

\subsubsection{Theoretical Properties}
\label{section:nondeterministic:algorithm:properties}

In this section we report some theoretical properties of our algorithm. At first we focus on $l$. We show that when the topology is fully connected the patroller receives the largest expected utility with any $l \geq \overline{l} = 0$. Subsequently, we show that, in arbitrary settings, the value $\overline{l}$ such that the patroller's expected utility is maximum is larger than 1 and therefore the Markovian equilibrium strategy we calculate is not the best one for the patroller. Finally, we show that, when searching for a non-deterministic equilibrium strategy, we cannot reduce the graph $G$ to a graph $G'$ where all the non-target vertices are removed as we have done in Section~3.

We start by discussing $\overline{l}$. We show that when the environment has a fully connected topology then $\overline{l}=0$. We state the following theorem.


\begin{thm}
For a fully connected topology, $\overline{l}=0$.\label{thm:fully}
\end{thm}
\emph{Proof sketch.} We prove that, in an environment with fully connected topology, our algorithm produces the same leader-follower equilibrium when $l=0$ and $l=1$. We consider the basic case with three vertices denoted by $\{1,2,3\}$, $d(i)=2$ for any vertex $i$ where agents are risk neutral. The proof in the general case with more complex patrolling settings and with $l>1$ is a generalization and we omit it. Suppose $l=1$.
Consider the bilinear programming problem of Section~\ref{subsez:programming} in which the best response of the intruder is supposed to be $\textit{enter-when}(2,1)$. The objective function can be written as $\sum_{i\neq2} v_{\mathbf{p}}(i)\cdot (1-p)+\sum_i v_{\mathbf{p}}(i)\cdot p$ where $p=\alpha_{1,2}+\alpha_{1,1}\alpha_{1,2}+\alpha_{1,3}\alpha_{3,2}$ is the probability to capture the intruder in vertex $2$. Since $\sum_{i} v_{\mathbf{p}}(i)\geq \sum_{i\neq2} v_{\mathbf{p}}(i)$, the maximization of the objective function can be reduced to the maximization of $p$. We denote by $EU_i(j)$ agent $i$'s expected utility from making action $j$. The constraints are (\emph{a}) $EU_{\mathbf{i}}(\textit{enter-when}(2,1))\geq EU_{\mathbf{i}}(\textit{enter-when}(2,2))$ and $EU_{\mathbf{i}}(\textit{enter-when}(2,1))$ $\geq$ $EU_{\mathbf{i}}(\textit{enter-when}(2,3))$, and (\emph{b}) $EU_{\mathbf{i}}(\textit{enter-when}(2,1))$ $\geq$ $EU_{\mathbf{i}}(\textit{enter-when}(i,j))$ for $i\in\{1,3\}$ and for $j\in\{1,2,3\}$. Consider the first constraint of (\emph{a}). It can be written as (with the same reduction used for the objective function): $\alpha_{1,2}+\alpha_{1,1}\alpha_{1,2}+\alpha_{1,3}\alpha_{3,2}\leq \alpha_{2,2}+\alpha_{2,1}\alpha_{1,2}+\alpha_{2,3}\alpha_{3,2}$. The second constraint of (\emph{a}) can be written analogously. Since the objective of the patroller is to maximize $\alpha_{1,2}+\alpha_{1,1}\alpha_{1,2}+\alpha_{1,3}\alpha_{3,2}$, we have that the maximum is when either $\alpha_{1,2}=\alpha_{2,2}=\alpha_{3,2}=0$ or $\alpha_{i,j}=\alpha_{k,j}$ for all $i,j,k$. The first option is not possible, since it prescribes that the patroller never patrols vertex $2$ knowing that the best response of the intruder is to enter vertex $2$. Thus, the second option holds. Under this option, the mathematical problem reduces to the one with $l=0$ and therefore they admit the same solution. In the situation in which constraints (\emph{b}) are more strict than constraints (\emph{a}), the corresponding problem with $l=0$ results unfeasible and there is an action of the intruder such that the utility expected by the patroller is larger than that expected when $\textit{enter-when}(2,1)$ (against the assumption that this is the best response for the intruder).\hfill$\Box$

With other topologies, it is generally  $\overline{l}\geq 1$. To see it, let us consider, the setting of Figure~\ref{F:graforunningexample} with the penetration times reported in Figure~\ref{F:graforidotto}(b). Such a patrolling problem admits a deterministic equilibrium strategy as showed in Section~3 and therefore the patroller can preserve all the value, forcing the intruder to make $\textit{stay-out}$. Now, suppose to apply the mathematical programming formulation of Section~4.2.3 to such problem. Being that formulation based on the assumption that $l=1$, the probability that the intruder will be captured when it makes, e.g., $\textit{enter-when}(06,23)$ is strictly lower than 1. Indeed,  the values $\alpha_{11,i}$ with $i\in\{06,12,18\}$ are strictly positive to assure that the patroller can cover all the targets. Then, by Markov chains, it follows that the probability that the patroller reaches vertex 06 starting from vertex 23 within 9 turns is strictly lower than 1. We can always find a strictly positive value of the penalty to be captured $\epsilon$ such that, when the patroller follows the Markovian strategy, the intruder strictly prefers to attack a target rather than not to attack. Since the intruder will attack and the probability of being captured is strictly lower than 1, the utility expected by the patroller from following the Markovian strategy (with $l=1$) will be strictly lower than that expected utility from following the deterministic equilibrium strategy (for which $l$ is indefinitely large). As this example shows, in general $\overline{l}>1$.

Now, we show that, in looking for a non-deterministic equilibrium patrolling strategy, we cannot study a patrolling setting considering only the targets and neglecting other vertices that connect them. We state the following proposition.

\begin{prop}
A non-deterministic patrolling strategy $\sigma_{\mathbf{p}}$ that is defined only on the space of the targets and prescribes that the patroller moves between targets along the shortest paths may be not optimal for the patroller.
\end{prop}
To show that this proposition holds, consider again our running example depicted in Figure~\ref{F:graforunningexample}. Suppose that $\sigma_{\mathbf{p}}$ is defined only on the targets, i.e., $\alpha_{i,j}$ is defined only when $i,j\in T$, and that the patroller moves between targets along the shortest paths. Since the intruder can attack a target both when the patroller is in a target and when it is moving along the paths, it can be easily observed that the patroller looses expected utility with respect to the situation where $\sigma_{\mathbf{p}}$ is defined on all the vertices. Indeed, the intruder can wait observing the actions of the patroller and deciding the turn in which to enter. Suppose that the intruder attacks vertex 18 after the patroller moved in vertex 21 from vertex 18. Since the patroller's strategy is defined only on the targets, it means that the patroller is going either 08 or 14. In both these two cases, the patroller will spend 6 turns. Then the patroller could came back to 18, but it needs more than 2 turns. Thus, if the intruder attacks 18 after the patroller moved in 21 from 18, then the intruder will surely have success in its attack. This does not happen when $\sigma_{\mathbf{p}}$ is defined on all the vertices $V_r$.

\subsection{Experimental Evaluation} \label{section:nondeterministic:experimental}

In this section we discuss the performance achieved by our algorithm in computing a non-deterministic equilibrium strategy. Differently from the experimental evaluation of the algorithm for computing a deterministic patrolling strategy presented in Section~\ref{S:deterministic:experiments}, the patrolling settings we tested here are not randomly generated, but have been carefully handcrafted to highlight some characteristics of our approach. Since the algorithm to find the non-deterministic strategy is much more computationally expensive than that employed in the deterministic case, the resolution of a large number of randomly generated instances would be computationally intractable, introducing difficulties in performing a good experimental evaluation. Furthermore, our main objective in this phase is not only to assess the absolute performance achieved by our algorithm, but also to evaluate the relative improvements that the reduction based on dominance of actions and the possibility to exploit a strictly competitive formulation could introduce. For these reasons, we conducted experiments on a number of patrolling settings, built with the aim of reproducing common situations that are likely to be found in real world applications. We assume agents to be risk neutral. For each setting we computed the non-deterministic equilibrium strategy with and without the reduction of dominated actions. Moreover, we also evaluated the performance when the setting allows one to compute the strategy using a strictly competitive formulation. In order to perform this last set of experiments, we modified functions $v_{\mathbf{p}}$ and $v_{\mathbf{i}}$ in the patrolling settings for which the requirements described in Section~\ref{section:nondeterministic:algorithm:strictlycompetitive} are not satisfied. All the algorithms have been coded in C and the optimization problems have been formulated in the AMPL~\cite{ampl} syntax and solved with the SNOPT~\cite{snopt} solver. Tests were conducted on a Linux (2.6.24 kernel) computer with a DUAL QUAD CORE Intel XEON 2.33 GHz CPU, 8 GB RAM, and 4 MB cache

\begin{figure}[htbp]
\centering
\begin{scriptsize}
\begin{tabular}{cc}
\resizebox*{0.5\textwidth}{!}{\includegraphics{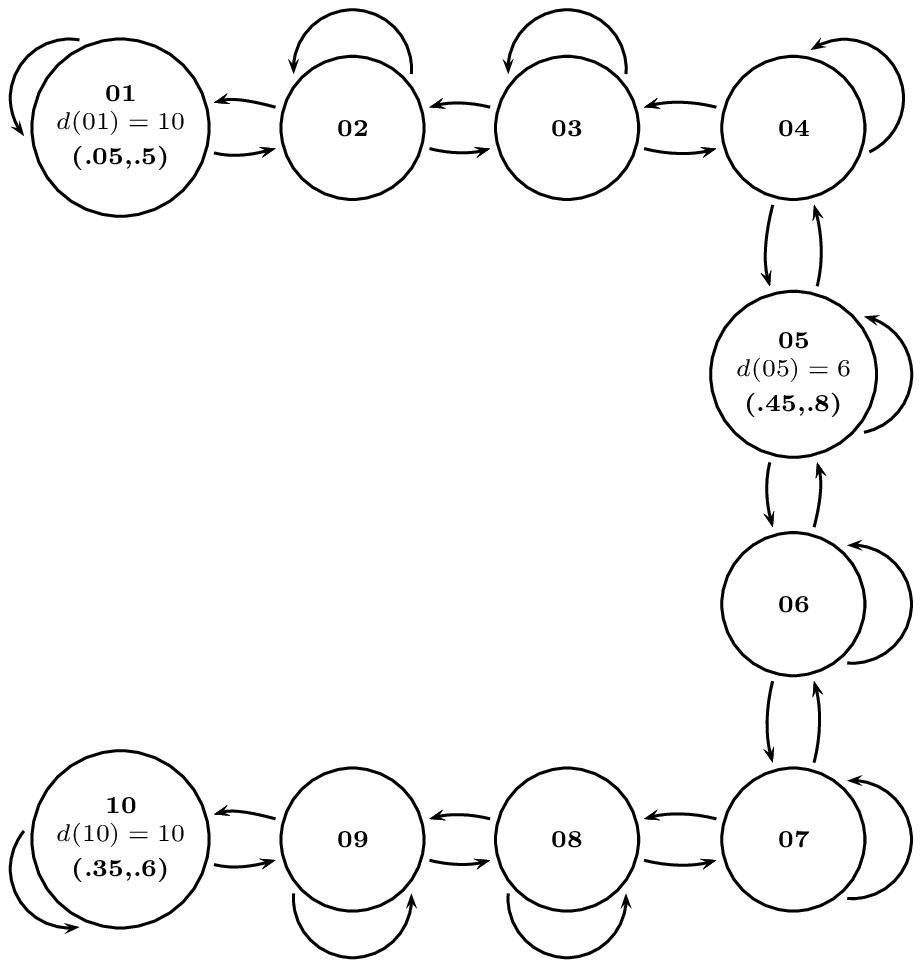}}   &
\raisebox{4cm}{
\begin{tabular}{|l|c|c|c|}
\hline
 & complete & reduced $\mathbf{p},\mathbf{i}$ & sc \\ \hline
total time & $9~\minute~14~\second$ & $28~\second$ & $3.5~\second$ \\ \hline
opt. prob. & 90 & 16 & 1 \\ \hline
average time & $5~\second$ & $5~\second$ & $-$ \\ \hline
max time & $9~\second$ & $10~\second$ & $-$ \\ \hline
min time & $<1~\second$ & $3~\second$ & $-$ \\ \hline
std dev. & $1~\second$ & $2~\second$ & $-$ \\ \hline
\end{tabular}
}
\end{tabular}
\end{scriptsize}
\caption{Results for a corridor-like setting with $10$ vertices.}
    \label{T:results.corridor}
\end{figure}

\begin{figure}[htbp]
\centering
\begin{scriptsize}
\begin{tabular}{cc}
\resizebox*{0.5\textwidth}{!}{\includegraphics{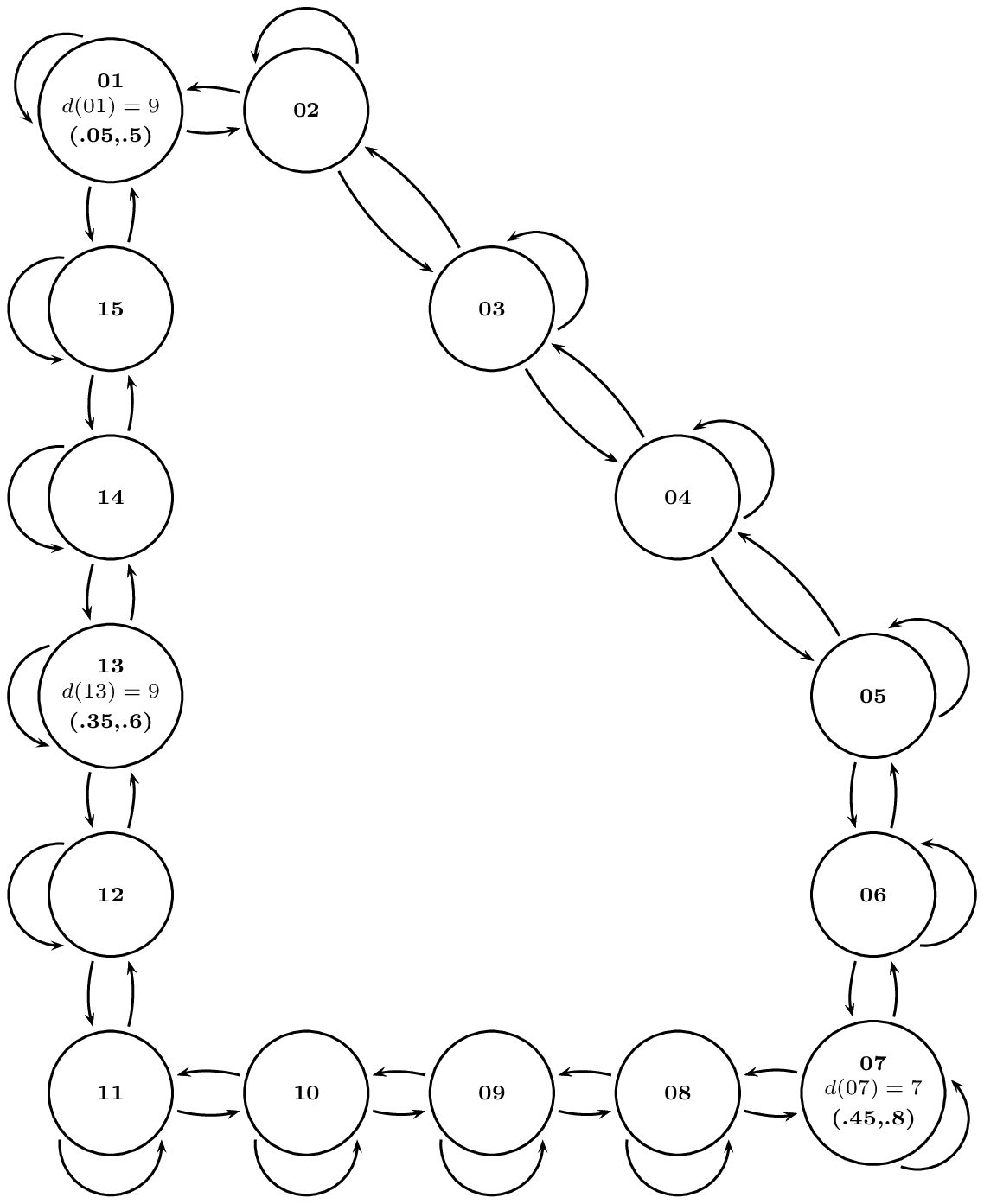}}   &
\raisebox{4cm}{
\begin{tabular}{|l|c|c|c|}
\hline
 & complete & reduced $\mathbf{p},\mathbf{i}$ & sc \\ \hline
total time & $3~\textnormal{h}~38~\minute$ & $11~\minute~49~\second$ & $49~\second$ \\ \hline
opt. prob. & 240 & 32 & 1 \\ \hline
average time & $57~\second$ & $44~\second$ & $-$ \\ \hline
max time & $1~\minute~28~\second$ & $95~\second$ & $-$ \\ \hline
min time & $28~\second$ & $10~\second$ & $-$ \\ \hline
std dev. & $10~\second$ & $16~\second$ & $-$ \\ \hline
\end{tabular}
}
\end{tabular}
\end{scriptsize}
\caption{Results for a ring-like setting with $15$ vertices.}
\label{T:results.ring}
\end{figure}

\begin{figure}[htbp]
\centering
\begin{scriptsize}
\begin{tabular}{cc}
\resizebox*{0.35\textwidth}{!}{\includegraphics{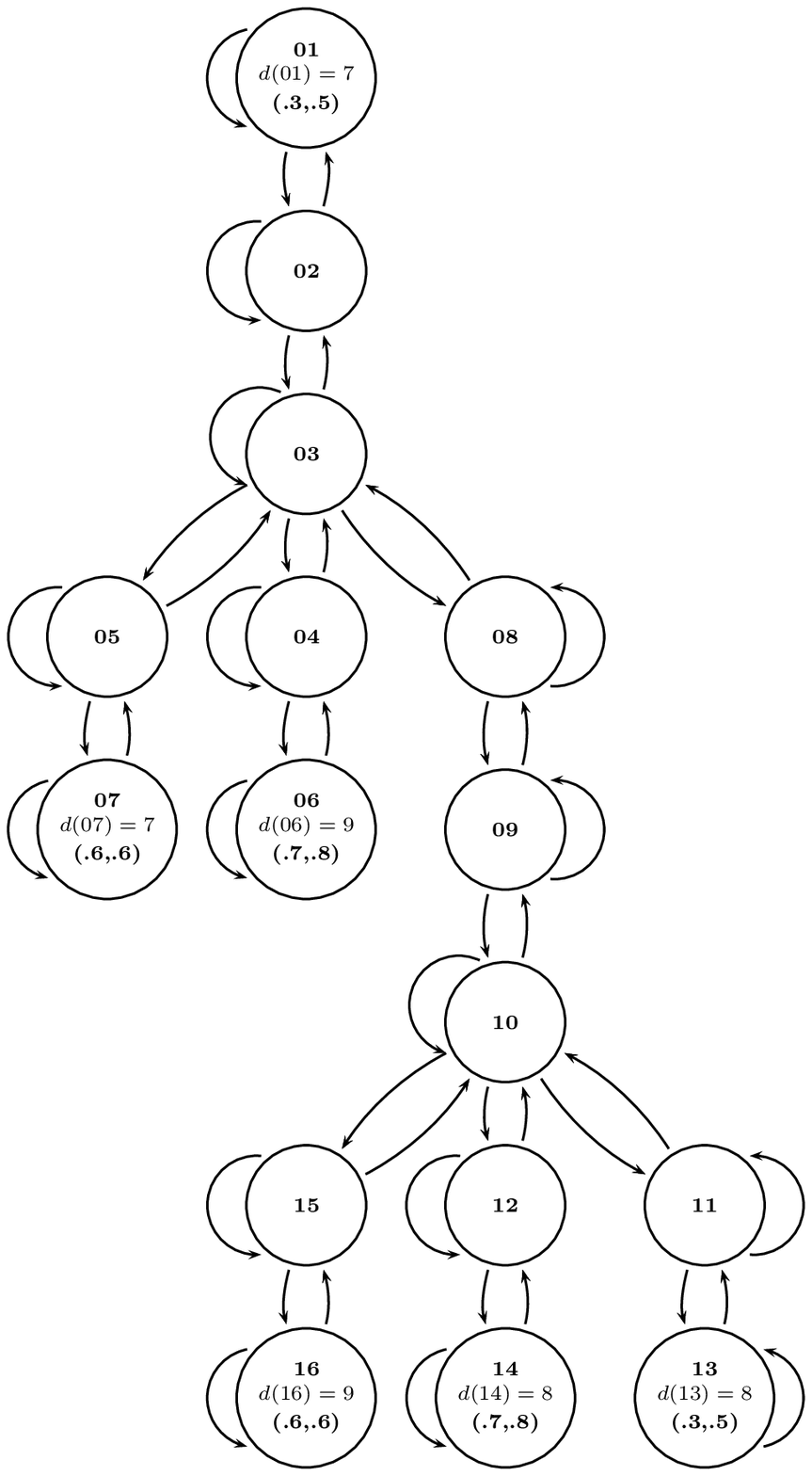}}   &
\raisebox{4cm}{
\begin{tabular}{|l|c|c|c|}
\hline
 & complete & reduced $\mathbf{p},\mathbf{i}$ & sc \\ \hline
total time & $35~\textnormal{h}~11~\minute$ & $27~\minute~14~\second$ & $1~\minute~3~\second$\\ \hline
opt. prob. & 240 & 32 & 1 \\ \hline
average time & $8~\minute~12~\second$ & $51~\second$ & $-$ \\ \hline
max time & $4~\textnormal{h}~22~\minute$ & $3~\minute~41~\second$ & $-$  \\ \hline
min time & $1~\second$ & $34~\second$ & $-$ \\ \hline
std dev. & $36~\minute~53~\second$ & $31~\second$ & $-$ \\ \hline
\end{tabular}
}
\end{tabular}
\end{scriptsize}
\caption{Results for a tree-like setting with $16$ vertices.}
\label{T:results.tree}
\end{figure}

\begin{figure}[htbp]
\centering
\begin{scriptsize}
\begin{tabular}{cc}
\resizebox*{0.4\textwidth}{!}{\includegraphics[width=0.7\textwidth]{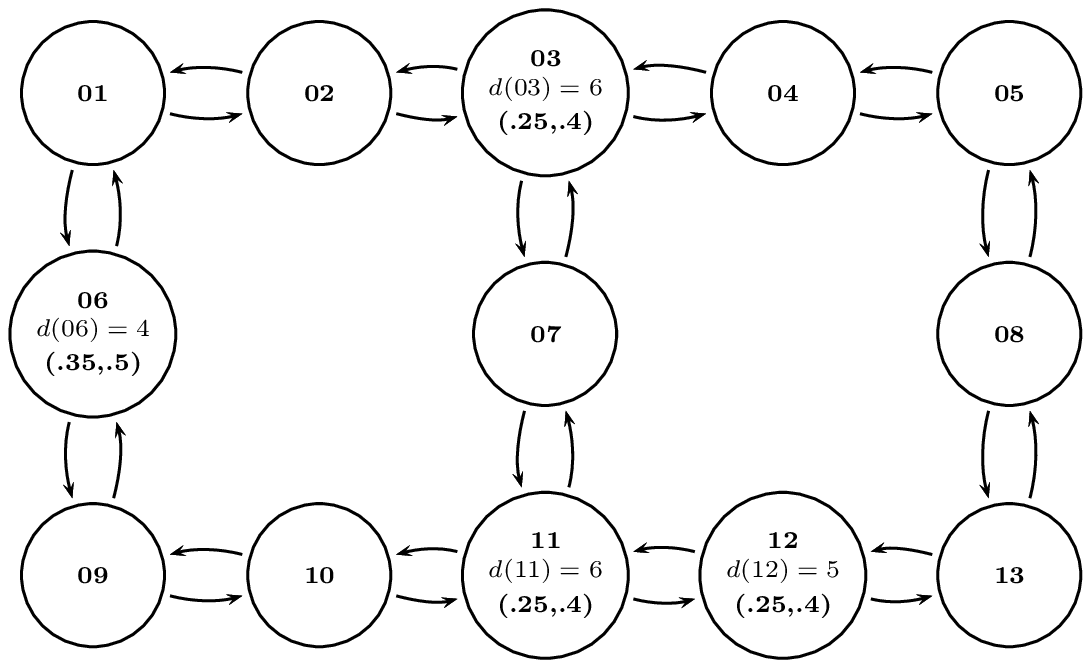}}   &
\raisebox{1.5cm}{
\begin{tabular}{|l|c|c|c|}
\hline
 & complete & reduced $\mathbf{p},\mathbf{i}$ & sc \\ \hline
total time & $6~\textnormal{h}~9~\minute$ & $26~\minute~49~\second$ & $6~\second$ \\ \hline
opt. prob. & 156 & 16 & 1 \\ \hline
average time & $1~\minute~25~\second$ & $1~\minute~9~\second$ & $-$  \\ \hline
max time & $5~\minute~1~\second$ & $2~\minute~31~\second$ & $-$  \\ \hline
min time & $1~\second$ & $46~\second$ & $-$   \\ \hline
std dev. & $30~\second$ & $21~\second$ & $-$  \\ \hline
\end{tabular}
}
\end{tabular}
\end{scriptsize}
\caption{Results for a eight-like setting with $13$ vertices.}
\label{T:results.papero}
\end{figure}

\begin{figure}[htbp]
\centering
\begin{scriptsize}
\begin{tabular}{cc}
\resizebox*{0.35\textwidth}{!}{\includegraphics[width=0.6\textwidth]{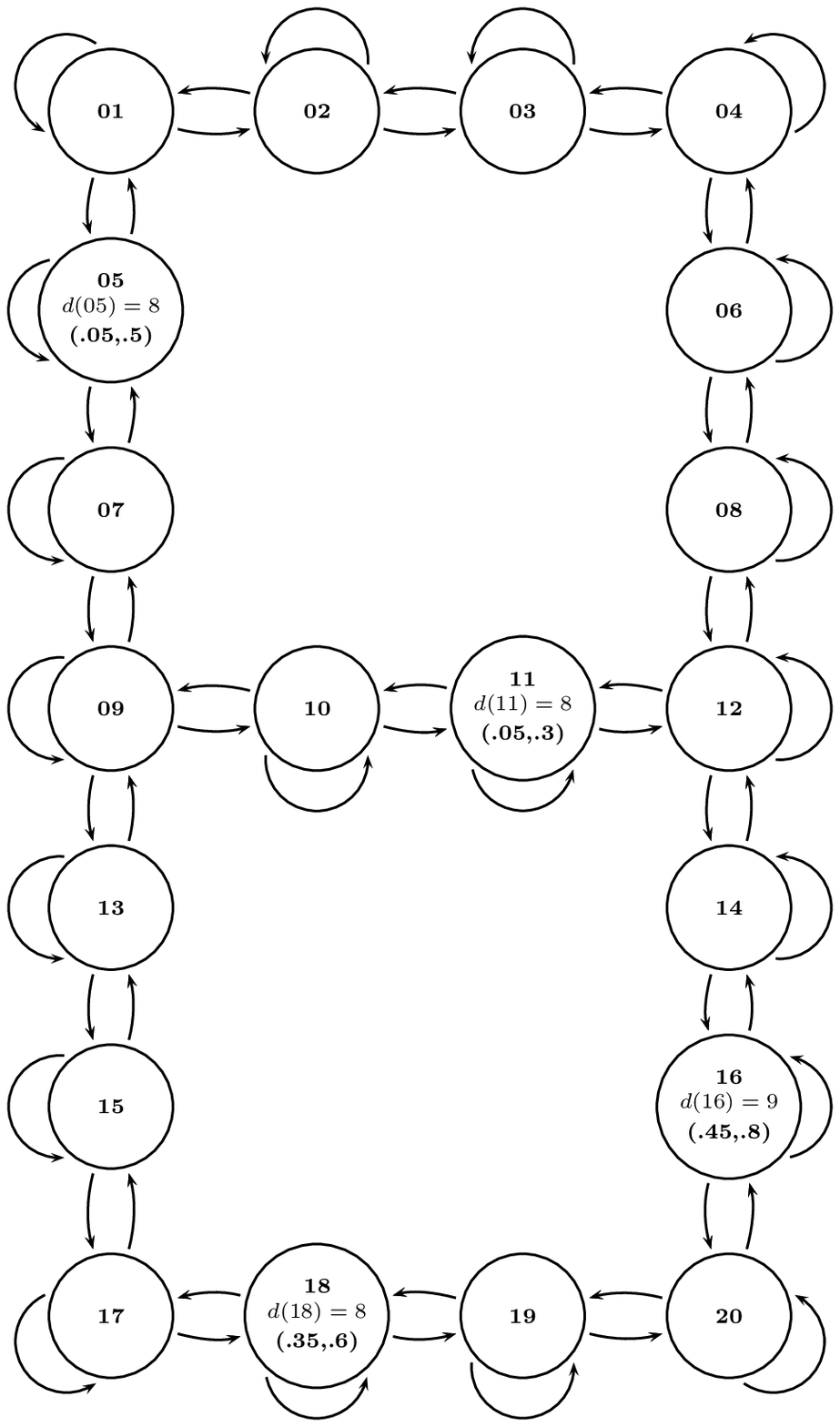}}   &
\raisebox{4cm}{
\begin{tabular}{|l|c|c|c|}
\hline
 & complete & reduced $\mathbf{p},\mathbf{i}$ & sc \\ \hline
total time & $46~\textnormal{h}~52~\minute$ & $1~\textnormal{h}~53~\minute$ & $4~\minute~45~\second$ \\ \hline
opt. prob. & 380 & 28 & 1 \\ \hline
average time & $6~\minute~59~\second$ & $4~\minute~1~\second$ & $-$ \\ \hline
max time & $2~\textnormal{h}~54~\minute$ & $9~\minute~51~\second$ & $-$  \\ \hline
min time & $2~\minute~38~\second$ & $1~\minute~42~\second$ & $-$  \\ \hline
std dev. & $8~\minute~29~\second$ & $1~\minute~43~\second$ & $-$ \\ \hline
\end{tabular}
}
\end{tabular}
\end{scriptsize}
\caption{Results for a eight-like setting with $20$ vertices.}
\label{T:results.eight}
\end{figure}

In order to have a fair evaluation, the considered patrolling settings are such that no equilibrium patrolling strategy inducing \emph{stay-out} as the intruder's best response could be found. In other words, to find the equilibrium patrolling strategy, all the three steps composing the algorithm for non-deterministic patrolling strategies need to be executed. We do not report the computational time spent for the removal of dominated actions, since it is negligible, being in all the settings shorter than one second. In Table~\ref{T:results.running}, we report the experimental results related to our running example of Figure~\ref{F:graforunningexample}. We consider four different configurations: `complete' refers to the basic case in which no dominated action is removed,  `reduced $\mathbf{p}$' refers to the case in which only patroller's dominated actions are removed, `reduced $\mathbf{p},\mathbf{i}$' refers to the case in which all the dominated actions are removed, and `sc' refers to the strictly competitive setting. We report the following data: `total time' is the computational time, `opt. prob.' is the number of optimization problems to be solved, `average time' is the average computational times to solve a single optimization problem, `max time' and `min time' are the largest and the shortest, respectively, computational time to solve a single optimization problem, `std dev.' is the standard deviation. We report in Figures~\ref{T:results.corridor}, \ref{T:results.ring}, \ref{T:results.tree}, \ref{T:results.papero}, and~\ref{T:results.eight} the graph representations of other patrolling settings and the corresponding experimental results. In these cases, `reduced $\mathbf{p}$' is omitted, since no patroller's action is dominated.

\begin{table}[htbp]
\centering
\begin{scriptsize}
\begin{tabular}{|l|c|c|c|c|}
\hline
 & complete & reduced $\mathbf{p}$ & reduced $\mathbf{p},\mathbf{i}$ & sc \\ \hline
total time & $>48~\textnormal{h}$ & $31~\textnormal{h}~21~\minute~25~\second$  & $29~\minute~13~\second$ &  $48~\second$ \\ \hline
opt. prob. & 812 & 240 & 30 & 1 \\ \hline
average time & - & $7~\minute~54~\second$ & $58~\second$ & $-$ \\ \hline
max time & - & $10~\textnormal{h}~4~\minute~52~\second$ & $1~\minute~49~\second$ & $-$ \\ \hline
min time & - & $46~\second$ & $8~\second$ & $-$ \\ \hline
std dev. & - & $51~\minute~55~\second$ & $3~\minute~16~\second$ & $-$ \\ \hline
\end{tabular}
\end{scriptsize}
\caption{Results for the running example of Figure~\ref{F:graforunningexample}.}
   \label{T:results.running}
\end{table}

As it can be seen for all the settings, when no reduction is applied, the computational time needed to find the solution is very large. Indeed, a considerable number of optimization problems have to be solved, most of which are not necessary, being associated to intruder's dominated actions. Moreover, each optimization problem is also characterized by many constraints and variables. Results show a remarkable improvement in performance when the reduction of the patrolling setting based on dominated actions is applied before computing the solution. An average $96\%$ reduction of the computational time was observed over all the settings for the non-strictly competitive case. This significant decreasing in the total computational time is directly related to the smaller number of optimization problems to be solved. However, a lower average time for the single problem has been obtained too. If we consider that the time needed to perform the reduction of dominated actions is negligible, such a pre-processing phase becomes of paramount importance for making our approach applicable, especially in complex patrolling settings.

Finally, in all the settings the best performance is achieved when using the strictly competitive formulation. Such a significant improvement is obviously due to the fact that only a single (and reduced) optimization problem needs to be solved. Despite this kind of formulation can be exploited only in specific cases, it constitutes a promising direction for the resolution of very complex patrolling settings.

\section{Extending the Framework}
\label{S:extensions} 

In this section, we discuss how our framework can be extended to capture some more realistic aspects and to improve its efficiency in finding a patrolling strategy. First, we enrich the expressiveness of the framework (Section~\ref{ss:refinement}) and, then, we propose some improvements for the algorithms (Section~\ref{ss:extensionimprovement}).

\subsection{Enriching Framework Expressiveness}
\label{ss:refinement}

We discuss how the framework expressiveness can be improved capturing the following situations. We introduce uncertainty over the intruder's valuations (Section~\ref{ss:uncertainvaluation}) and over the intruder's penetration times (Section~\ref{ss:uncertainpenetration}), and we enrich the patroller sensing capabilities (Section~\ref{ss:uncertainsensing}). Then we extend the intruder's movement model (Section~\ref{ss:intrudermovement}), we introduce delay on intruder's entering (Section~\ref{ss:uncertaindelay}), we introduce intruder's partial observability over patroller's actions (Section~\ref{ss:partialobservation}), and we consider the situation in which there are multiple patrolling robots (Section~\ref{ss:multiplerobot}). For each extension, we discuss its impact on the model and on the computation of deterministic and non-deterministic equilibrium strategies.

\subsubsection{Uncertainty over Intruder's Valuations}
\label{ss:uncertainvaluation}
We consider the situation wherein the patroller does not know with certainty the intruder's valuations $v_{\mathbf{i}}$s over the targets. 
\begin{description}
\item[Model] According to Harsanyi's transformation~\cite{fudenberg1991}, a game with uncertain information is cast to a game with imperfect information. More precisely, each player can be of different types, each one with different values of the payoffs, and the players do not perfectly observe the actual type of the opponents, e.g., see~\cite{paruchuri2008}. In our patrolling game, the intruder could be of different types $\theta_j\in \Theta_{\mathbf{i}}$, each one with specific valuations $v_{\mathbf{i}}(i,\theta_{j})$ over target $i$ and penalty $\epsilon_j$ for being captured. Each intruder's type $\theta_j$ is assigned a probability $\omega_j$. The intruder strategy $\sigma_{\mathbf{i}}$ must define a strategy for each type of intruder.
\item[Deterministic Equilibrium Computation] No extension is required. Indeed, the algorithm for the computation of the deterministic equilibrium strategy described in Section~3 does not depend on the intruder's valuations.
\item[Non-Deterministic Equilibrium Computation] The step of the algorithm that computes the dominated actions does not require any extension, since it does not depend on the intruder's valuations. Considering the non-strictly competitive case, the bilinear mathematical programming formulation must be extended as follows. We need to compute the best patroller's strategy for each profile of actions of intruder's types, e.g., with two intruder's types for each $\langle \textit{enter-when}(i_1, h_1), \textit{enter-when}(i_2,h_2)\rangle$ where $i_j$, $h_j$ corresponds to the $j$-th type. This requires the introduction of constraints for each intruder's type in the mathematical programming problem. As a results, both the number of constraints and the number of mathematical programming problems to be solved increase with a power of the number of types $|\Theta_{\mathbf{i}}|$. This makes the exact resolution of real-world settings impractical and pushes for finding approximated algorithms. Let us consider the strictly competitive case. By definition, the intruder's valuations can be any under the constraint that they are strictly competitive with respect to the patroller's ones. In the case all the possible intruder's types have valuations that are strictly competitive with respect to the patroller's ones, we can ignore the intruder's types and we study the game as a classical strictly competitive game where the intruder's valuations are those of a possible type. This makes the exact resolution of significant complex settings affordable.
\end{description}

\subsubsection{Uncertainty over Intruder's Penetration Times}
\label{ss:uncertainpenetration}
We consider the situation wherein penetration times of the intruder are uncertain due to the fact that the environment is not perfectly predictable or modeled.
\begin{description}
\item[Model] Each target $t$'s penetration time $d(t)$ is described by a discrete probability distribution, e.g., 
\[d(06)=\begin{cases}8&\textnormal{with probability } 0.3 \\ 9&\textnormal{with probability }0.4 \\ 10&\textnormal{with probability }0.3\end{cases}.\]
\item[Deterministic Equilibrium Computation] For each target, we consider the shortest possible penetration time according to the given probability distribution (8, in the example above). Beyond that, the algorithm works as in the basic case, but considering such penetration time.
\item[Non-Deterministic Equilibrium Computation] For each target, we consider the largest possible penetration time according to the given probability distribution (10, in the example above). The computation of the dominated actions works as in the basic case, but using such penetration time. The mathematical programming formulations, for both the non-strictly competitive and the strictly competitive cases, must be extended as follows. The value of $w$ in constraints~(\ref{con:feasibility5}) is in the range $\{2,\ldots,\max(d(t))\}$, where $\max(d(t))$ is the largest possible penetration time for target $t$ according to the given probability distribution. The computation of the intruder's expected utility in constraints~(\ref{con:feasibility6}) must we weighted with the probabilities of the possible values of $d(t)$. The number of constraints and the number of optimization problems to be solved is the same as in the basic case.
\end{description}

\subsubsection{Augmented Patroller's Sensing Capabilities}
\label{ss:uncertainsensing}
We consider the situation wherein patroller's sensing capabilities are augmented. A more detailed description of this situation can be found in~\cite{BasilicoGattiRossiCIG2009}.
\begin{description}
\item[Model] The model has been already described in Section~\ref{subsection:problemstatement} and uses a generic function $S(\cdot,\cdot)$ to model the sensing capabilities of the patroller.
\item[Deterministic Equilibrium Computation] This kind of strategy can be computed only if we consider sensing with probability of one. Formally, we require that $S(i,j)=0$ or $S(i,j)=1$ for all vertices $i$ and $j$. The generation of the reduced graph $G'$ must be modified to consider $S$. In practice, $G'$ is composed also of additional vertices that are on the boundary of the sensing range with respect to a target, i.e., the farthest vertices from which the patroller can sense the target. The algorithm can then be easily extended to work on this graph.
\item[Non-Deterministic Equilibrium Computation] In the computation of the dominated actions, action $\textit{enter-when}(i,j)$ is dominated by action $\textit{enter-when}(i,k)$ if it results to be dominated according to the algorithm provided in Section~\ref{S:nondeterministic:solving:domination} and $S(j,i)\geq S(j,k)$. The mathematical programming problems (for both the non-strictly competitive and the strictly competitive cases) can be easily modified to capture probabilistic augmented sensing capabilities. More precisely, the right term of constraints~(\ref{con:feasibility5}) and~(\ref{con:feasibility6}) must be multiplied by $1-S(i,j)$. This does not increase the number of constraints and the number of optimization problems to be solved in the non-strictly competitive case and, as shown in~\cite{BasilicoGattiRossiCIG2009}, no significant additional computational time is needed to address this extension. 
\end{description}

\subsubsection{Refinements on Intruder's Movement Model}
\label{ss:intrudermovement}
We extend the movement model of the intruder. A more detailed description of this extension can be found in~\cite{BasilicoGattiRossiCeppiAmigoniIAT2009}.
\begin{description}
\item[Model] We enrich the model by introducing \emph{access areas}, which are special vertices, constraining the intruder to move along paths connecting access areas to targets, and allowing the patroller to capture the intruder also along these paths.  As it is often the case in pursuit-evasion, we assume that the intruder can move infinitely fast. Therefore, covering a path takes only one turn, regardless of its length. The intruder's actions are now of the form $\textit{enter-when}(p,i)$ where $p$ is a path connecting an access area to a target. By employing search trees the minimal set of paths to be considered can be computed.
\item[Deterministic Equilibrium Computation] The algorithm is exactly the same we described in Section~3 except for the special case in which all the possible intruder's paths share a common vertex. In this case, the optimal strategy for the patroller is obviously to stay forever in such vertex.
\item[Non-Deterministic Equilibrium Computation] In the computation of the dominated actions, action $\textit{enter-when}(p,j)$ is dominated by action $\textit{enter-when}(p,k)$ if it results to be dominated according to the algorithm provided in Section~\ref{S:nondeterministic:solving:domination} and $k$ is not adjacent to any vertex of $p$. The mathematical programming problems (both non-strictly competitive and strictly competitive) can be easily modified to capture intruder's movement along paths. More precisely, the right term of constraints~(\ref{con:feasibility5}) must consider that all the vertices belonging to $p$ are sensed when $w=1$. This does not increase the number of constraints and the number of optimization problems to be solved in the non-strictly competitive case and, as shown in~\cite{BasilicoGattiRossiCeppiAmigoniIAT2009}, no significant additional computational time is required to address this extension.
\end{description}

\subsubsection{Delay on Intruder's Entering}
\label{ss:uncertaindelay}
We consider the situation wherein there is a delay between the turn at which the intruder decides to enter and the turn at which it actually enters. A more detailed description of this situation can be found in~\cite{BasilicoGattiRossiCIG2009}.
\begin{description}
\item[Model] We introduce a delay $D \in \mathbb{N}^{+}$ between the turn at which the intruder decides to enter a target $t$ and the turn at which it actually enters $t$. A probability distribution can be defined over this delay.
\item[Deterministic Equilibrium Computation] No modification is required with respect to the algorithm described in Section~3.
\item[Non-Deterministic Equilibrium Computation] In the computation of the dominated actions, we need to add delay $D$ to the penetration times of each target. The mathematical programming problems (both non-strictly competitive and strictly competitive) must be modified as follows. Constraints~(\ref{con:feasibility5}) must be rewritten considering that the range of $w$ is $\{D+1,\ldots, D+d(t)\}$. We need to introduce constraints similar to constraints~(\ref{con:feasibility6}) with $w$ in $\{2,\ldots, D\}$. These constraints enable to derive an estimate on the position of the patroller at the turn in which the intruder enters. This extension makes the range of $w$ longer, introducing additional constraints; however, as shown in~\cite{BasilicoGattiRossiCIG2009}, no significant additional computational time is needed to address this extension.
\end{description}

\subsubsection{Intruder's Partial Observation over Patroller's Actions}
\label{ss:partialobservation}
We consider the situation wherein the intruder, when it decides to enter, can partially observe the position of the patroller. A more detailed description of this situation can be found in~\cite{BasilicoGattiRossiCeppiAmigoniIAT2009}.
\begin{description}
\item[Model] As in Section~\ref{ss:intrudermovement}, we enrich the model by introducing access areas and constraining the intruder to enter the patrolling setting through these access areas. We assign each access area a view over the vertices, i.e., the set of vertices that can be observed from such access area. Since the intruder cannot perfectly observe the position of the patroller when it enters and can enter when the patroller is out of the view of the intruder, we need to redefine the intruder's action space. Specifically, we define a \emph{state} $s=\langle j, k \rangle$ where $j$ is a vertex and $k$ is an integer denoting time. The meaning of $s$ is: at the last observation, $k$ turns before, the patroller was in vertex $j$. Thus, the intruder's actions are $\textit{enter-when}(i,s)$, where $i$ is a target and $s$ is a state.
\item[Deterministic Equilibrium Computation] No modification is required with respect to the algorithm described in Section~3.
\item[Non-Deterministic Equilibrium Computation] We need to redefine the computation of the dominated actions to take into account the possibility that the intruder enters when the patroller is not observable. In principle, the intruder's actions are infinite, being infinite the values that $k$ can assume in $s$. Anyway, it can be shown that only a finite number of actions are non dominated. That is, we can safely consider a finite number of values for $k$. The mathematical programming problems must be modified in a way similar to that of Section~\ref{ss:uncertaindelay}. Indeed, to compute the patroller's best strategy when the intruder makes $\textit{enter-when}(i,s)$ with $s = \langle j,k \rangle$ and $k>0$, we need to estimate the position of the patroller. This estimation is based on the patroller's strategy. As shown in~\cite{BasilicoGattiRossiCeppiAmigoniIAT2009}, addressing this extension can require significant additional computational time.
\end{description}

\subsubsection{Multiple Patrolling Robots}
\label{ss:multiplerobot}
We consider the situation wherein there are multiple patrolling robots. A more detailed description of this situation can be found in~\cite{AmigoniBasilicoGattiICRA2009}.
\begin{description}
\item[Model] Multiple robots can be easily captured in the model by exploiting patroller's augmented sensing capabilities. Let us assume that the environment is a perimeter and that the robots move together in a synchronized fashion (i.e., at each turn they all move clockwise or counterclockwise, as in~\cite{GalKaminkaICRA2008}). We can define a fictitious single robot with an appropriate $S(i,j)$ such that it senses all the vertices that the real robots would sense. In the general case, we can model the position of the robots as a tuple and their actions as a joint action. The intruder's actions are of the form $\textit{enter-when}(i,\langle j_1,\ldots, j_g\rangle)$ where $j_r$ is the position of the $r$-th robot.
\item[Deterministic Equilibrium Computation] The algorithm described in Section~3 must be modified as follows. The solution is a tuple of strategies, one for each specific robot. That is, the algorithm produces a tuple of cycles. At each step of the search tree, the algorithm builds upon the current partial solution by adding a target to the strategy of each robot. This target will be different for each robot.
\item[Non-Deterministic Equilibrium Computation] The determination of the intruder's dominated actions must consider the presence of multiple robots. Specifically, the algorithm presented in Section~\ref{S:nondeterministic:solving:domination} must be repeated for each patrolling robot. An action is dominated if it is dominated with respect to all the robots. The number of constraints in the mathematical programming problems (for both the non-strictly competitive and the strictly competitive cases) increases since the intruder's space of actions is larger. The number of optimization problems to be solved in the non-strictly competitive case increases exponentially with $g$ (the number of robots). The considerations on computational time are similar to those reported in Section~\ref{ss:uncertainvaluation}.
\end{description}

\subsection{Improving Algorithms}
\label{ss:extensionimprovement}

In this section, we discuss how our solving algorithm can be extended to situations in which the optimal patrolling strategy does not cover all the targets (Section~\ref{ss:nonfullycoveraged}) and how approximated solutions can be found (Section~\ref{ss:epsilon}).
For each extension, we discuss its impact on the computation of deterministic (when applicable) and non-deterministic equilibria.

\subsubsection{Computing Non-Full Coverage Strategies}
\label{ss:nonfullycoveraged}
We consider the situation in which the optimal strategy does not cover all the targets and we cannot introduce additional patrolling robots.
\begin{description}
\item[Deterministic Equilibrium Computation] In this case, the aim is to determine the subset of targets to be patrolled such that the patroller's utility is the largest. This can be accomplished by removing iteratively the targets in increasing order of patroller's utility and searching at each iteration for a deterministic equilibrium strategy as prescribed in Section~3. The algorithm stops when it finds a subset of targets that admits a deterministic equilibrium strategy.
\item[Non-Deterministic Equilibrium Computation] No extension is required for the computation of the intruder's dominated actions. The computation of the non-deterministic equilibrium strategy can be accomplished iteratively similarly to the computation of the deterministic equilibrium strategy. More precisely, initially all the targets are considered. If the leader-follower equilibrium strategy is such that some target $t$ is not covered, then we apply our algorithm to a new patrolling problem where target $t$ is removed.
\end{description}

\subsubsection{Searching for $\epsilon$-Equilibria}
\label{ss:epsilon}
We consider the problem of computing approximate solutions for the mathematical programming problems that find the non-deterministic equilibrium.
\begin{description}
\item[Non-Deterministic Equilibrium Computation] The hardness of the problem of searching for an exact leader-follower equilibrium can be relaxed by looking at approximated equilibria, called $\epsilon$\emph{-equilibria}. A strategy is an $\epsilon$-equilibrium if each player cannot increase its expected utility more than $\epsilon$ by making an off-equilibrium action. The literature provides several algorithms to compute these equilibria, but only few of them are applicable for leader-follower settings, see, e.g.,~\cite{Letchford}. Thus, this remains an open problem.
\end{description}


\section{Conclusions}
\label{S:conclusions}

In this paper we have presented a formal framework for addressing the problem of finding optimal patrolling strategies for a mobile robot moving in an environment to prevent intrusions. The proposed approach is based on the idea of modeling a patrolling situation as a game, played by the patroller and the intruder, and of studying its equilibria to derive the optimal patrolling strategy. The approach is more general than other game theoretical approaches presented in literature, since it deals with environments with arbitrary topology and with arbitrary preferences for the agents. Starting from the formal definitions of a patrolling setting and of the associated game, the main contributions of the paper have been algorithms for finding equilibrium patrolling strategies both in a deterministic and in a non-deterministic form. In the first case, the equilibrium patrolling strategy is a fixed path that, when followed by the patrolling robot, makes attempting an attack not rational for an intruder. In the second case, the equilibrium patrolling strategy is a set of probabilities for moving between different positions that, when followed by the patrolling robots, maximizes its expected utility. Both the algorithms have been analytically studied and experimentally validated to assess their properties and efficiency.

Several avenues for future work have been outlined in Section~\ref{S:extensions}. Some preliminary work has been started along a number of these directions, but most problems remain open. In addition, a fundamental aspect that is being addressed is the application of the theoretical framework presented in this paper to real robots~\cite{armor, tambeAAMAS2009} and, more generally, the throwing of a bridge between studies of patrolling in the mobile robot community and those in the theoretical community.

\section*{Acknowledgments}
We are glad to thank Edoardo Amaldi, Alessandro Busi, Sofia Ceppi, Stefano Coniglio, and Claudio Iuliano for their fruitful suggestions and collaboration.

\bibliographystyle{elsarticle-num}
\bibliography{AIJPatrolling}

\end{document}